\documentclass[9pt,twocolumn,twoside,dvipsnames]{pnas-new}

\templatetype{pnasresearcharticle} 

\newbox{\orcidlogo}
\sbox{\orcidlogo}{\large\includegraphics[height=1.8ex]{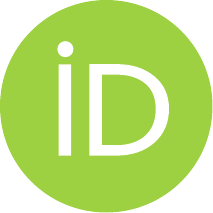}}
\newcommand{\orcid}[1]{\href{https://orcid.org/#1}{\usebox{\orcidlogo}\,}}

\title{Metagenomic Analysis using Phylogenetic Placement -- A Review of the First Decade}

\author[a,*]{Lucas Czech \orcid{0000-0002-1340-9644}}
\author[b,c]{Alexandros Stamatakis \orcid{0000-0003-0353-0691}}
\author[d]{Micah Dunthorn \orcid{0000-0003-1376-4109}}
\author[*]{Pierre Barbera \orcid{0000-0002-3437-150X}}

\affil[a]{Department of Plant Biology, Carnegie Institution for Science, Stanford, USA}
\affil[b]{Computational Molecular Evolution Group, Heidelberg Institute for Theoretical Studies, Heidelberg, Germany}
\affil[c]{Institute for Theoretical Informatics, Karlsruhe Institute of Technology, Karlsruhe, Germany}
\affil[d]{Natural History Museum, University of Oslo, Oslo, Norway}
\affil[*]{To whom correspondence should be addressed. E-mail: \href{mailto:lczech@carnegiescience.edu}{\color{black}{lczech@carnegiescience.edu}} and \href{mailto:pierre@barbera-bio.info}{\color{black}{pierre@barbera-bio.info}}}

\newcommand{\mainauthor}{Czech \it{et al.}}
\leadauthor{\mainauthor}

\setboolean{displaywatermark}{false}

\usepackage{tcolorbox}

\usepackage{listings}
\usepackage{lastpage}
\usepackage{verbatimbox}

\usepackage{caption}
\usepackage[labelformat=simple]{subcaption}

\definecolor{MyLightSteelBlue1}{RGB}{218,236,255}
\definecolor{MyLightSteelBlue4}{RGB}{100,113,159}

\usepackage[binary-units=true]{siunitx}

\usepackage{tocloft}
\tocloftpagestyle{fancy}
\usepackage{titlesec}
\renewcommand{\thesubsection}{\arabic{section}.\arabic{subsection}}

\newcommand{\figref}[1]{Figure~\ref{#1}}
\newcommand{\tabref}[1]{Table~\ref{#1}}
\newcommand{\secref}[1]{Section~``\nameref{#1}''}

\newcommand\toolname{\textsc}

\newcommand{\codeline}[1]{\texttt{\lstinline|#1|}}

\newcommand\fileformat{\texttt}

\newcommand\entityname{\textit}

\keywords{
}

\begin{abstract}
Phylogenetic placement refers to a family of tools and methods to analyze, visualize, and interpret
the tsunami of metagenomic sequencing data generated by high-throughput sequencing.
Compared to alternative (e.\,g., similarity-based) methods,
it puts metabarcoding sequences into a phylogenetic context using a set of known reference sequences and taking evolutionary history into account. Thereby, one can increase the accuracy of metagenomic surveys and eliminate the requirement for having exact or close matches with existing sequence databases.
\\
Phylogenetic placement constitutes a valuable analysis tool \emph{per se},
but also entails a plethora of downstream tools to interpret its results.
A common use case is to analyze species communities obtained from metagenomic sequencing,
for example via taxonomic assignment, diversity quantification, sample comparison,
and identification of correlations with environmental variables.
\\
In this review, we provide an overview over the methods developed during the first ten years.
In particular, the goals of this review are
(i)~to motivate the usage of phylogenetic placement and illustrate some of its use cases,
(ii)~to outline the full workflow, from raw sequences to publishable figures, including best practices,
(iii)~to introduce the most common tools and methods and their capabilities,
(iv)~to point out common placement pitfalls and misconceptions,
(v)~to showcase typical placement-based analyses,
and how they can help to analyze, visualize, and interpret phylogenetic placement data.
\\
\vspace{-0.5em}
\\
\textbf{Contact:}
\href{mailto:lczech@carnegiescience.edu}{lczech@carnegiescience.edu} and
\href{mailto:pierre@barbera-bio.info}{pierre@barbera-bio.info}.
\end{abstract}

\begin{document}

\verticaladjustment{-2pt}

\maketitle
\thispagestyle{firststyle}

\ifthenelse{\boolean{shortarticle}}{\ifthenelse{\boolean{singlecolumn}}{\abscontentformatted}{\abscontent}}{}


\vspace{-1.5em}

\tableofcontents

\vspace{-0.2em}


\section{Introduction}
\label{sec:Introduction}

Advances in sequencing technologies enable the broad sequencing of genetic material in environmental samples \citep{Edwards2013,Sunagawa2013a}, for instance, from water \cite{Karsenti2011,Giner2016,LacoursireRoussel2016}, soil \cite{Dupont2016,Mahe2017}, and air \citep{Clare2022}, which is known as environmental DNA \citep[eDNA, ][]{Deiner2017,Ruppert2019}, or from the human body \cite{Huttenhower2012,Methe2012,Matsen2015,Wang2015} and other sources \cite{Hanson2016,Gohli2019,Lorimer2019,ElRakaiby2019}.
Crucially, this enables the ecological survey of a community of organisms in their immediate environment (i.\,e., \emph{in situ}), and allows to directly study the genetic composition of species communities (from viruses to megafauna);
a field known as metagenomics \citep{Thomas2012,Oulas2015,Escobar-Zepeda2015,Lindgreen2016}.

Metagenomic data typically stem from so-called \emph{High-Throughput Sequencing} \citep[HTS,][]{Pettersson2009,Reuter2015,Goodwin2016} technologies,
such as \emph{Next Generation Sequencing} \citep[NGS,][]{Logares2012,Mardis2013},
as well as later generations \citep{Pareek2011,Niedringhaus2011,Mignardi2014,Heather2016,Mardis2017}.
For a sample of biological material, these technologies typically produce thousands to millions or even billions of short genetic sequences (also called ``reads'') with a length of some hundred base pairs length each.
Over the past decades, decreasing costs and increasing throughput of sequencing technologies
have caused an exponential growth in sequencing data \citep{Muir2016}, which has now passed the peta-scale barrier \citep{Katz2022}.

A major analysis step in metagenomic studies is to characterize the reads obtained from an environment by means of comparison to \emph{reference sequences} of known species \citep{Desai2012}.
A straight-forward way to accomplish this is to quantify the similarity between the reads and reference sequences.
We obtain an indication of possible novelty if the sequence similarity to known species is low \cite{Temperton2012,Peabody2015}.
However, such approaches do not provide the user with the evolutionary context of the read, and have been found to incorrectly identify sequences \citep{Koski2001,Clemente2011,Mahe2017}.

Instead, general phylogenetic methods can be used directly to classify and characterize the reads, providing highly accurate and information-rich results \citep{Brady2009,Segata2012,Truong2015,Jamy2019a,Beghini2021}.
However, trying to resolve the phylogenetic relationships between millions of short reads and the given reference sequences represents a significant computational challenge.
Furthermore, as most phylogenetic methods require an \emph{alignment} of sequences, metagenomic data can often not be used directly, as whole-genome reference data might not be available or computationally intractable.
Instead, specific \emph{marker genes} can be targeted (or filtered from the metagenomic data), which are genetic regions that are well-suited for differentiating between species \cite{Ren2016}.
The use of marker genes to identify species is called \emph{DNA (meta-)\hspace{0pt}barcoding} \cite{Hebert2003,Savolainen2005,Kress2008,Deiner2017}; see \secref{sec:PhylogeneticPlacement:sub:TypesOfSequences} for details.

A powerful and increasingly popular class of methods to identify and analyze diverse (meta-)genomic (barcode) data is the so-called \emph{phylogenetic placement} (or \emph{evolutionary placement})
of genetic sequences onto a given fixed phylogenetic \emph{reference tree}.
By placing unknown, anonymous sequences (in this context called \emph{query sequences}) into the evolutionary context of a tree,
these methods allow for the taxonomic assignment of the sequences
\citep[i.\,e., the association of genomic reads to existing species, for example  ][]{Jamy2019a,Auladell2019,Hleap2021}.
Moreover, they can also provide information on the evolutionary relationships
between these query sequences and the reference species/sequences, and thus go beyond simple species identification.
Phylogenetic placement has found applications in a variety of situations, such as data cleaning and retention \citep{Mahe2017}, inference of new clades \citep{Dunthorn2014,Bass2018a}, estimation of ecological profiles \citep{Keck2018}, identification of low-coverage genomes of viral strains \citep{Muhlemann2020a}, phylogenetic analysis of viruses such as SARS-CoV-2 \citep{Morel2020sars,Turakhia2021},
and in clinical studies of microbial diseases \citep{Srinivasan2012}.


When analyzing the resulting data, there are two complementary interpretations of phylogenetic placement:
(1) as a set of individual sequences, placed with respect to the reference phylogeny, e.\,g., for taxonomic assignment, phylo-geographic tracing, or even possible clinical relevance;
(2) as a combined distribution of sequences on the tree, characterizing the sampled environment at a given point in time or space to examine the composition of a species community as a whole,
for instance as a means of sample ordination and visualization, and association with environmental variables.

In this review, we provide an overview of existing methods to conduct phylogenetic placement,
as well as post-analysis methods for visualization and knowledge inference from placement data.
We also discuss some practical aspects, such as common pitfalls and misconceptions,
as well as caveats and limitations of these methods.
We mainly refer to metagenomic input data (or more accurately, metabarcoding data, see below for details) as it represents the most common use case, but also highlight some alternative use cases where phylogenetic placement is employed for other types of sequence data.


\begin{tcolorbox}[
    title={Glossary and Abbreviations}, fonttitle=\sffamily\bfseries,
    float=htb, boxsep=5pt, left=0pt, right=0pt, top=0pt, bottom=0pt,
    colframe=MyLightSteelBlue4, colback=MyLightSteelBlue1
]
    \setlength{\parskip}{1ex plus 0.5ex minus 0.2ex}
    \sffamily
    \small


    \textbf{Likelihood Weight Ratio (LWR).} The probability (confidence) that a QS is placed
    onto a particular branch (i.\,e., a single Placement Location). 

    \textbf{Maximum Likelihood (ML).} A statistical framework to estimate the parameters of a probability distribution.



    \textbf{Phylogenetic Placement.} A family of methods to place a set of QSs onto the branches of an RT,
    by mapping each QS to one or several most likely Placement Locations on the tree.

    \textbf{Placement Location.} An individual location (branch and position along the branch) onto which a specific QS has been placed;
    often annotated with a probability score (LWR) whose sum over all branches is 1 for that QS.


    \textbf{Query Sequence (QS).} A single sequence to be placed into the RT.
    Typically, this is a short read or amplicon obtained via metabarcoding or metagenomics.

    \textbf{Reference Alignment (RA).} The underlying multiple sequence alignment (MSA),
    based on a set of RSs, that is used in ML-based phylogenetic placement and was used to infer the RT.

    \textbf{Reference Sequence (RS).} A typically high-quality sequence of a species or strain that is used as reference to compare the QSs
    against. Used to compute the RA and infer the RT.

    \textbf{Reference Tree (RT).} The (bifurcating) phylogenetic tree used as a scaffold to place the QSs into, mostly inferred via ML methods.
\end{tcolorbox}

\section{Phylogenetic Placement}
\label{sec:PhylogeneticPlacement}


\subsection{Overview and Terminology}
\label{sec:PhylogeneticPlacement:sub:Overview}


The modern approach to phylogenetic tree inference is based on molecular sequence data,
and uses stochastic models of sequence evolution \citep{Arenas2015} to infer the tree topology
and its branch lengths \citep{Felsenstein2004,Yang2006}.
Note that the computational cost to infer the optimal tree under the given optimality criterion grows super-exponentially in the number of sequences \cite{Felsenstein2004}.
In addition, large trees comprising more than a couple of hundred sequences are often cumbersome to visualize, rendering the approach challenging for current (e.\,g., metagenomic) large datasets.
Furthermore, the lack of phylogenetic signal contained in the short reads of most HTS technology usually does not suffice for a robust tree inference \citep{Bininda-Emonds2001,Moret2002b,VonMering2007,Dunthorn2014}.
Hence, \emph{phylogenetic placement} emerged from the demand
to obtain phylogenetic information about sequence sets
that are too large in number and too short in length
to infer comprehensive phylogenetic trees \citep{Matsen2010,Berger2011}.
In a metagenomic context, a set of sequences obtained from an environment such as water, soil, or the human body,
is here called a \emph{sample}.
This is often the data that we intend to place, and might have further metadata associated with it, e.\,g., environmental factors/variables such as temperature or geo-locations where the sample was taken.

Generally, the input of a phylogenetic placement analysis is a phylogenetic \emph{Reference Tree} (RT) consisting of sequences spanning the genetic diversity that is expected in the sequences to be placed into the tree.
The tree can be rooted or unrooted; in the latter case however, a ``virtual'' root (or top-level trifurcation) is used in the computation as a fixed point of reference \cite{Czech2017}.
Then, for a single sequence (e.\,g., a short read), in this context called a \emph{Query Sequence} (QS),
the goal of phylogenetic placement is to determine the branches of the RT
to which the QS is most closely evolutionarily related.
Note that the RT is kept fixed, that is, the QSs are not inserted as new branches into the tree,
but rather ``mapped'' onto its branches. Hence, the phylogenetic relationships {\em between} individual QSs are not resolved.

This is the key insight that makes it possible to efficiently compute the placement of large numbers of QSs.
By only determining the evolutionary relationship between the sequences of the RT and each individual QS,
the process can be efficiently parallelized, and the 
required processing time scales linearly in the number of QS.
Furthermore, this allows us to consider multiple branches as potential \emph{Placement Locations} for a given QS,
representing uncertainty in the placement,
often expressed as a probability (or confidence) of the QS being placed on that branch.
This uncertainty might result from weak phylogenetic signal,
or might indicate some other issue with the data, as explained later.
In maximum-likelihood (ML) based placement (see \secref{sec:PhylogeneticPlacement:sub:GeneralPurpose:par:ML} for details),
these probabilities are computed as the \emph{Likelihood Weight Ratio} (LWR)
resulting from the evaluation of placing the QS attached to an additional (hypothetical) branch into the tree.
Hence, for historic reasons, the probability of a placement location (one QS placed on a specific branch)
is often called its LWR, and for a given QS, the sum of LWRs over all branches is 1
(equivalent to the total probability).
See \tabref{tab:PlacementTools} for an overview of different placement tools,
and which of the aforementioned quantities they can compute.

\begin{figure*}[!htb]
    \centering
    \includegraphics[width=\linewidth]{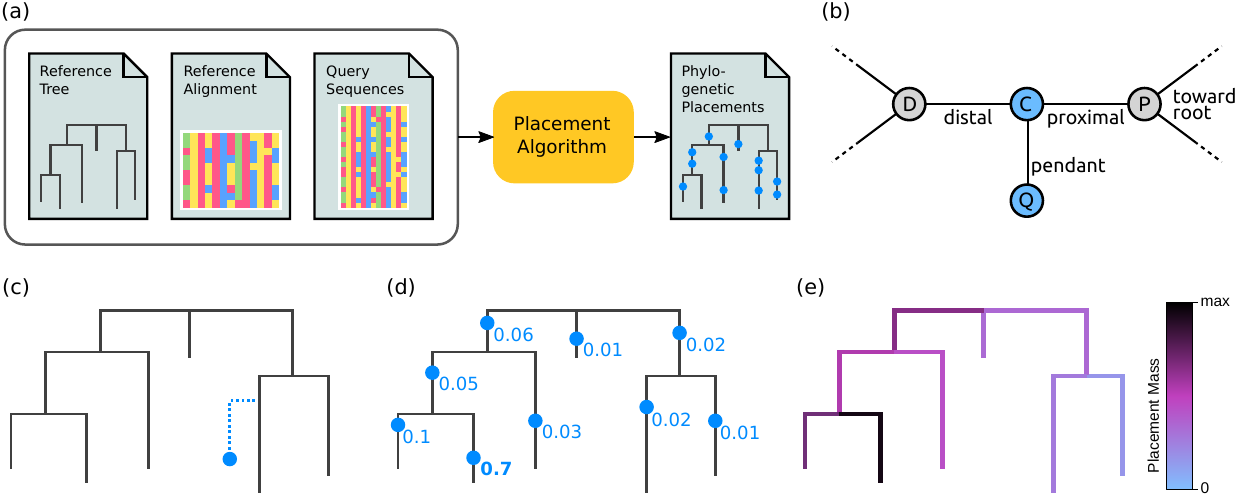}
    \begin{subfigure}{0pt}
        \phantomcaption
        \label{fig:Placement:sub:Pipeline}
    \end{subfigure}
    \begin{subfigure}{0pt}
        \phantomcaption
        \label{fig:Placement:sub:TinyTree}
    \end{subfigure}
    \begin{subfigure}{0pt}
        \phantomcaption
        \label{fig:Placement:sub:PlacementLocation}
    \end{subfigure}
    \begin{subfigure}{0pt}
        \phantomcaption
        \label{fig:Placement:sub:LWR}
    \end{subfigure}
    \begin{subfigure}{0pt}
        \phantomcaption
        \label{fig:Placement:sub:Distribution}
    \end{subfigure}
    \vspace*{-1em}
    \caption{
        \textbf{Overview of phylogenetic placement.}
        Here, we show the typical process, focused on ML-based placement. For the sake of simplicity, we here omit heuristics
        and other algorithmic improvements.
        Alignment-free placement works conceptually in an analogous way, but does not compute tree likelihoods.
        \hspace{1em}\subref{fig:Placement:sub:Pipeline}
        Pipeline and data flow.
        The input to phylogenetic placement are the Reference Tree (RT) and its corresponding Reference Alignment (RA), as well as the set of Query Sequences (QSs) that we are interested in.
        The placement algorithm computes potential placement locations of a QS on the branches of the RT, for each QS in the input.
        \hspace{1em}\subref{fig:Placement:sub:TinyTree}
        Terminology. 
        The nodes {\sffamily D} and {\sffamily P} belong to the Reference Tree (RT).
        When placing a Query Sequence (QS), the branch between these nodes 
        is split into two parts by a temporary new node {\sffamily C},
        which serves as the attachment point for another temporary new node {\sffamily Q} that represents the QS.
        Note that these two new nodes are only conceptually inserted into the RT -- they represent the mapping of the QS onto that branch.
        The \emph{pendant} branch leads to {\sffamily Q}.
        The original branch is split into the \emph{proximal} branch, which leads towards the (possibly virtual) root of the RT,
        and the \emph{distal} branch, which leads away from the root.
        \hspace{1em}\subref{fig:Placement:sub:PlacementLocation}
        A single QS is placed onto a single branch (that is, one placement location).
        Vertical distances symbolize branch lengths.
        Note that the QS is located at a certain position along its Reference Tree branch
        (splitting that branch into distal and proximal parts), and has a (pendant) branch length of its own.
        At this step, ML-based placement computes the likelihood of the RT with the QS as a (temporary) extra branch.
        For one single QS, this step is then repeated at every branch of the tree.
        \hspace{1em}\subref{fig:Placement:sub:LWR}
        Once the likelihoods of placing the QS onto every branch have been computed,
        the Likelihood Weight Ratios (LWRs) for this QS are computed.
        They express the confidence of placing the QS onto each branch,
        and can be interpreted as a probability distribution of the QS across the tree
        (and hence sum to one across all branches).
        In the image, we omit pendant branch lengths for the sake of simplicity.
        \hspace{1em}\subref{fig:Placement:sub:Distribution}
        The process is repeated for every QS, yielding an LWR-weighted ``mapping'' of each QS to each branch.
        We can visualize this as a cumulative distribution across all QSs on the tree,
        coloring branches according to the total sum of the LWRs at that branch over all QS.
        See \figref{fig:Examine:sub:MassTree} for a real-world example of this.
    }
\label{fig:Placement}
\end{figure*}

In other words, phylogenetic placement can be thought of as an all-to-all mapping from QSs to branches of the RT,
with a probability for each placement location,
as shown in \figref{fig:Placement:sub:LWR} and \figref{fig:Placement:sub:Distribution}.
We can however also interpret each such placement location \textit{as if} it was an extra branch inserted into the RT,
as shown in \figref{fig:Placement:sub:TinyTree} and \figref{fig:Placement:sub:PlacementLocation}.
In particular, maximum likelihood placement makes use of its underlying evolutionary model
to also estimate the involved branch lengths that are altered through the insertion of a QS, see \figref{fig:Placement:sub:TinyTree} for details.
This interpretation highlights the aspect of each individual QS being part of the underlying phylogeny.
For example, this allows its taxonomic assignment to that clade of the reference tree
where the QS shows the highest accumulated placement probability, as explained later.






\paragraph{Misconceptions}
\label{sec:PhylogeneticPlacement:sub:Concept:par:Misconceptions}

In the existing literature, and from our experience in teaching the topic as well as supporting the users of our software,
some concepts of phylogenetic placement are not always well explained or understood.
Although we have introduced these concepts above already,
we briefly address two common misconceptions here, for clarity.

Firstly, a common misconception is that the tree is amended by the QSs, that is, that new branches are added to the RT,
and that the phylogenetic relationships of the QSs with each other are hence resolved.
This is not the case; instead, the RT is kept fixed, the QSs are only aligned against the reference alignment, but not against each other (in ML placement), and the QSs are mapped only to the existing branches in the RT.
This mapping \textit{can} however be interpreted ``as if'' the QS was a new terminal node (leaf or tip) of the tree,
usually inserted (or ``grafted'') into the branch with the most probable placement location,
which can be useful in some applications.

Secondly, a further common misconception is that a QS is only placed onto a single branch,
or that only the best (most likely) placement location is taken as the result for each placed QS.
Instead, each branch is seen as a potential placement location with a certain probability, which sum to one over the tree.
It can however be useful to reduce the placement distribution of a QS to only its most probable placement location.
Also, for practical reasons, typically not all locations are stored in the resulting file
(or even considered in the computation by application of heuristics),
as low probability locations can often be discarded to save storage space and downstream processing time;
see \secref{sec:PhylogeneticPlacement:sub:Concept:par:FileFormat} for details.
Lastly, some placement methods do only output a single best placement, see~\tabref{tab:PlacementTools}.


In summary, phylogenetic placement yields a distribution of potential locations of where a QS could be attached in the RT  -- but it does not extend the RT by the QS with an actual branch.


\paragraph{File Format}
\label{sec:PhylogeneticPlacement:sub:Concept:par:FileFormat}

\begin{figure}[!tb]
    \centering
    \includegraphics[width=\linewidth]{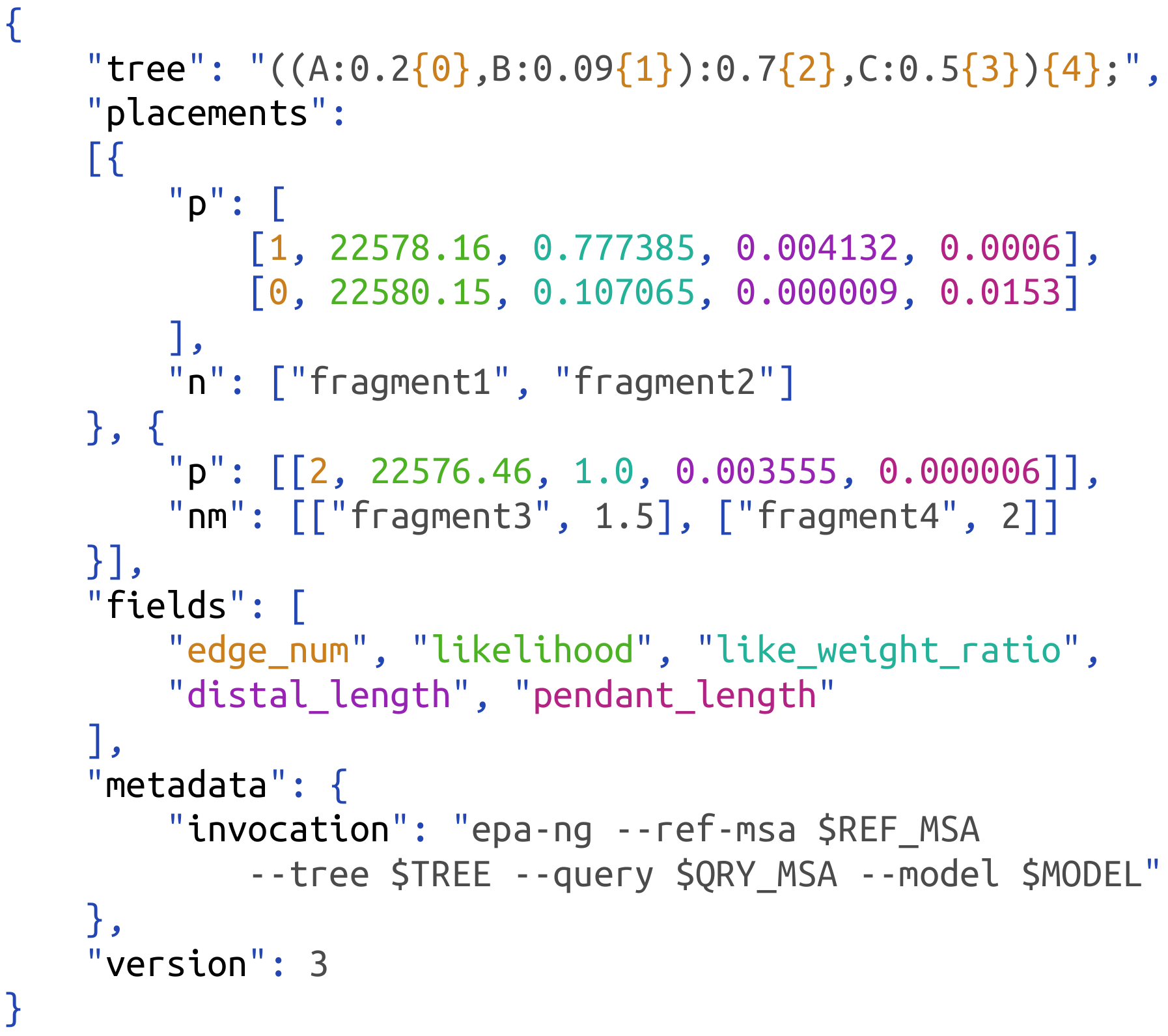}
    \vspace*{-1em}
    \caption{
        \textbf{Jplace format for phylogenetic placement.}
        The exemplary file consists of a reference \texttt{"tree"} in a custom Newick format that annotates edge numbers
        in curly brackets,
        followed by two \emph{pqueries}, which is the term for combined lists of sequence names and their placement locations.
        The first pquery contains two placement locations (\texttt{"p"}) for two query sequences (\texttt{"n"}),
        and the second contains a single location (\texttt{"p"}) for two other sequences including their multiplicities/abundances (\texttt{"nm"}).
        The order to interpret the values per location is given via the \texttt{"fields"} list, and highlighted by colors here;
        additional \texttt{"metadata"} and a \texttt{"version"} of the file format can be given.
        Example adapted from \citep{Matsen2012}.
    }
\label{fig:Jplace}
\end{figure}

Placement data is usually stored in the so-called \fileformat{jplace} format \citep{Matsen2012},
which is based on the \fileformat{json} format \citep{JsonStandard,JsonMemo}.
See \figref{fig:Jplace} for an example.
It uses a custom augmentation of the \fileformat{Newick} format \citep{Archie1986} to store the reference tree,
where each branch is additionally annotated by a unique edge number,
so that placement locations can easily refer to the branches.
For each QS (named via the list \verb|"n"|), the format then stores a set of possible placement locations (in the list \verb|"p"|), where each location is described by the values:
(1)~\verb|"edge_num"|, which identifies the branch of this placement location,
(2)~\verb|"likelihood"|, which is used by maximum likelihood based placement methods,
(3)~\verb|"like_weight_ratio"| (LWR),
which denotes the probability (or confidence) of this placement location for the given QS,
(4)~\verb|"distal_length"| and
(5)~\verb|"pendant_length"|, which are the branch lengths involved in the placement
of the QS for the given placement location; see~\figref{fig:Placement:sub:TinyTree} for an explanation of these lengths.

These five data fields are the standard fields of the \fileformat{jplace} format; further fields can be added as needed.
As noted above, typically not all placement locations for a given QS are stored in the file,
as low probability placements unnecessarily increase the file size without providing substantial information;
in that case, the sum of the stored LWR values might actually be smaller than~$1$.

The format furthermore allows for multiple names in the \verb|"n"| list,
as well as assigning a ``multiplicity'' to each such name (by using a list called \verb|"nm"| instead of \verb|"n"|).
For instance, this allows to only store the placement locations for identical reads once,
while keeping track of the original raw abundances of these reads or OTUs.
A pair of a \verb|"n"|/\verb|"nm"| list and a \verb|"p"| list is called a ``pquery'',
and describes a set of placement locations for one or more (identical) QSs.
This structure is then repeated for each QS that has been placed.

To our knowledge, the \toolname{genesis} library \citep{Czech2020} is the only general purpose toolkit for working with, and manipulating, placement data in \fileformat{jplace} format. It also incorporates many of the downstream visualization and analysis techniques we describe later on.
Some other tools that offer basic capability to work with \fileformat{jplace} files are \toolname{BoSSA} \citep{bossa}, \toolname{ggtree} \citep{Yu2017}, and \toolname{treeio} \citep{Wang2020}, all of which can read \fileformat{jplace} files for processing in R.

With the release of several placement tools that do not use the ML framework, see \secref{sec:PhylogeneticPlacement:sub:GeneralPurpose:par:DistanceBasedPlacement}, the \fileformat{jplace} file format \citep{Matsen2012} may require an update.
The standard is written currently (as of version 3) with placement properties such as branch lengths and likelihood scores in mind, which do not translate well to other types of placement algorithms (pers.~comm.~with S.~Mirarab, July 2020).
Furthermore, it might be helpful to support sample names, multiple samples per file, and additional per-sample or even per-query annotations and other metadata in the file format.
Being based on \fileformat{json}, this can already be achieved now by adding these entries ad-hoc, but would lack support by parsers if not properly standardized.



\subsection{Types of Query Sequences}
\label{sec:PhylogeneticPlacement:sub:TypesOfSequences}

In principle, any type of genetic sequence data can be subjected to placement, as long as the reference sequences span the genomic regions where the query sequences originate from.
Apart from the availability of suitable reference sequences used to construct a reference tree (see \secref{sec:PhylogeneticPlacement:sub:ReferenceSequencesAlignmentTree:par:SequenceSelection}),
the primary limiting factor is the extent to which a given placement tool supports the data.
Currently, the majority of placement tools supports nucleotide (DNA/RNA) and amino acid (protein) data.
Many placement methods require query reads to be aligned to the reference, i.\,e. they need to be homologs.


\paragraph{Metabarcoding and Amplicons}
\label{sec:PhylogeneticPlacement:sub:TypesOfSequences:par:Metabarcoding}

For the above reasons, a common approach to obtain sequences is \emph{metabarcoding} \cite{Hebert2003,Savolainen2005,Kress2008,Deiner2017}.
In metabarcoding, one or several \emph{marker} or \emph{barcoding} genes, such as 16S \cite{Weisburg1991}, 18S \cite{Meyer2010}, ITS, COI, etc. \cite{Woese1977,Woese1990,Ji2013,Sunagawa2013a} are typically chosen to compute the reference alignment, and appropriate primers are selected to enable metabarcode sequencing of the sample \citep{Deiner2017}.
A marker gene should be universally present in the studied organisms, and ideally should only occur once in the genome of each organism \citep{Nguyen2014,Dunthorn2014}, i.\,e., be single-copy.
In practice, marker genes often occur multiple times per genome, possibly requiring the need for copy number correction.
A marker gene should exhibit sufficient between-species variation to distinguish them from each other,
but show low within-species variation \cite{Kress2008}.
Using a metabarcoding approach has several advantages: it targets loci of interest and focuses the sequencing effort there (incidentally also limiting the size of the reference MSA), barcoding genes are typically well suited for phylogenetics (stable regions to aid alignment paired with variable regions to discriminate organisms), and the approach is generally cost-effective.
Such approaches use amplicon sequencing \citep{Peabody2015,Hugerth2017}, wherein only DNA originating from the targeted region is amplified using the Polymerase Chain Reaction \citep[PCR, ][]{Bartlett2003}, thus yielding the subsequent sequencing of any remaining DNA fragments from other regions highly improbable.
The resulting amplicon sequences have been shown to be well-suited for phylogenetic placement \citep{Mahe2017,Janssen2018}.

However, PCR-based amplifications are known to introduce biases in the abundance of the sequencing reads, as some fragments may be copied with a higher likelihood than others \cite{Logares2014,Morgan2010}.
Similarly, a further bias that skews abundance results exists as different organisms may have a different number of copies of the targeted gene, ranging from single copies to $15$ copies, depending on the organism \citep{Lee2009}.
Some methods exist that attempt to account for copy number bias \citep{Kembel2012,Angly2014,PereiraFlores2019} as well as for PCR amplification bias \citep{Love2016,Silverman2021}.

When an untargeted sequencing approach is chosen instead (such as shotgun metagenomic sequencing), using a broader scope for the reference sequences may be advisable, such as using whole genome data.
This might only be feasible for small genomes such as some viruses or mitochondrial DNA.
Alternatively, a sensible approach is to filter out any reads that did likely not originate from the genetic regions that constitute the reference alignment.
This can be achieved, for example, using \toolname{hmmsearch} from the \toolname{HMMER}-package \citep{Eddy1995,Eddy1998}, which allows the user to obtain a list of reads that have an alignment score above a given threshold.
Similarly, so-called \textsubscript{mi}tags \cite{Logares2014} represent a shotgun-based alternative to amplicon sequencing.

Recently, placement methods have emerged that do not require the alignment of query sequences to a reference, and some do not even require the references to be aligned against each other (see~\secref{sec:PhylogeneticPlacement:sub:GeneralPurpose:par:DistanceBasedPlacement}).
However, establishing that query reads and reference sequences are homologous is still necessary.


\paragraph{Sequencing Technologies}
\label{sec:PhylogeneticPlacement:sub:TypesOfSequences:par:Technologies}

A further consideration is the choice of sequencing technology, with the primary property being the length of the resulting sequencing reads.
So far, the vast majority of studies utilizing phylogenetic placement have relied on short-read sequencing technologies such as NGS, using by now well established protocols to perform broad low-cost sequencing \citep{vanDijk2014}.
However, this approach produces very short (150-400 nucleotide) reads, that typically only cover fragments of a reference gene. For universal single-copy markers, this can limit their applicability to phylogenetics due to the lower information content. However, the approach has been applied successfully to other types of data \citep{Piredda2021,Cardoni2022}.

More recent sequencing technologies, called third generation sequencing, or long-read sequencing (LRS), yield individual reads that cover entire genes, or even entire genomes \citep{Amarasinghe2020}.
While placement was originally developed for short read sequencing, longer read lengths typically increase the phylogenetic signal contained in reads, thus increasing the 
reliability of phylogenetic methods.
Indeed, such sequence data have been shown to overcome this fundamental hurdle to phylogenetically resolving the relationships between query sequences that originally gave rise to phylogenetic placement \cite{Jamy2019a}.

An emerging third way to obtain longer reads is to combine short reads into longer so-called Synthetic Long-Reads (SLRs), which have been used successfully to characterize metagenomes \citep{Sharon2015, Kuleshov2016} and which improve upon short-read metabarcoding approaches for taxonomic classification \citep{Jamy2019a,Ritter2020,Jeong2021}.


\paragraph{Clustering}
\label{sec:PhylogeneticPlacement:sub:TypesOfSequences:par:Clustering}

Once the wet-lab sequencing strategy has been determined, a user eventually obtains a (typically large) set of sequences.
After quality control, a potential next step is to consider if, and how, to cluster these raw sequences in order to reduce the amount of data that has to be processed, often at the cost of losing information.
Common choices include clustering by similarity threshold ($\geq97\%$) resulting in Operational Taxonomic Units \citep[OTUs,][]{Blaxter2005,Edgar2010,Fu2012,Rognes2016,Westcott2015}, more strictly based on single nucleotide differences resulting in Amplicon Sequencing Variants \citep[ASVs,][]{Callahan2016}, or more recent alternatives such as \toolname{SWARM} clustering \citep{swarm3}.
These methods are most commonly used for clustering reads from marker regions, and hence applicable in the placement context;
for a comprehensive review of clustering methods, see \cite{Zou2018}.

If possible, it is recommended to avoid clustering, in order to retain potential phylogenetic signal; this choice however also depends on study design and goals.
However, even if sequences are not clustered, we strongly recommend dereplication, that is, removal of exact (strict) duplicates of sequences, to avoid unnecessary redundant computations.
For the same reason, sequence dereplication is also useful when pooling the sequences from multiple samples together and placing the resulting set via a single placement run.
Tools that offer this capability include \toolname{USEARCH} \citep{Edgar2010}, and \toolname{VSEARCH} \citep{Rognes2016}, as well as the placement-specific \toolname{chunkify} command in \toolname{gappa} \citep{Czech2020}.


\paragraph{Outgroup Rooting}
\label{sec:PhylogeneticPlacement:sub:TypesOfSequences:par:OutgroupRooting}

Finally, an often overlooked source of query sequences are high-quality reference sequence databases.
Here, the use-case of placement shifts away from taxonomic assignment: instead such data can be used to attempt an outgroup rooting of an existing tree, using already classified sequences \citep{Hubert2014,Morel2020sars,Liede-Schumann2020}.
The result of placement, in this case, is a set of suggested branches on which to root the tree, including a probability estimate for each root placement onto each branch \citep{Liede-Schumann2020}.

\subsection{Reference Sequences, Alignment, and Tree}
\label{sec:PhylogeneticPlacement:sub:ReferenceSequencesAlignmentTree}

The phylogenetic reference tree (RT), inferred from a set of reference sequences (RSs) using their alignment (\emph{Reference Alignment}, RA),
is the foundation and scaffold for conducting phylogenetic placement.
Ideally, to avoid duplicating work, to ensure high quality,
and to provide stable points of reference for comparison between studies, suitable reference trees should be provided by the respective research/organismal communities.
First efforts for microbial eukaryotes are on their way \cite{Berney2017,DelCampo2018,Rajter2021,Rajter2021a},
although some of these are not designed explicitly for phylogenetic placements,
but more taxonomic groups will follow.
As such, references are however not yet available for all taxonomic groups, we here provide an overview of the process \citep[see also][for practical examples]{Mahe2017,Rajter2021}.


\paragraph{Sequence Selection}
\label{sec:PhylogeneticPlacement:sub:ReferenceSequencesAlignmentTree:par:SequenceSelection}

As phylogenetic placement cannot infer evolutionary relationships below the taxonomic level of the reference tree,
the first step is the selection of suitable RSs,
which should (i) cover the diversity that is expected in the query sequences (QSs), and
(ii) be well-established and representative for their respective clades to facilitate meaningful interpretation.
In order to capture unexpected diversity and potential outliers,
it can be advantageous to include a wider range of sequences as well \citep{Mahe2017},
or to run preliminary tests and filtering (placement- or similarity-based) with a broad reference 
to ensure that all diversity in the QSs is accounted for.

In many cases, the selection process is (unfortunately) labor-intense,
as it requires hand-selecting known sequences from reference databases such as
\toolname{SILVA} \citep{Pruesse2007,Quast2013,Yilmaz2014}, NCBI \citep{Sayers2009,Benson2009},
\toolname{GreenGenes} \citep{DeSantis2006,McDonald2012}, or \toolname{RDP} \citep{Cole2014,Wang2007}.
This manual process however also often provides the highest quality, and allows to optimally assemble the RSs for a given project.
See also \citep{Balvociute2017} for a comparison of these databases.

Important selection criteria are the number of sequences to be selected, as well as their diversity;
both of which depend on the study design and goals.
Generally, a number of RSs in the order of hundreds to a few thousands
has shown to provide enough coverage for most QS datasets,
while still being small enough to properly visualize their phylogeny and to conduct all necessary computations in reasonable time.
Often, it is sufficient to include a single species to represent a whole clade \citep{Rajter2021a}.
Depending on the types of downstream analyses,
it can be a disadvantage to select sequences that are too similar to each other
(i.\,e., closely related species, or different strains of the same species), as this can spread the placement distribution across nearby branches.
In other words, placements with similar probability in many branches are mostly a consequence of reference alignment regions for which large subtrees contain (almost) identical sequences.
This is however expected when conducting taxonomic assignment at species or below-species level, and the reference should be built with the targeted taxonomic resolution in mind.

On the other hand, if the QSs contain enough phylogenetic signal (e.\,g., when using long reads, whole genome data, or when the target gene has sufficient variability),
including multiple representatives of a taxonomic group might allow to obtain more finely resolved placements.
For example, in short genomes such as HIV or arthropod mitochondria, where mutations are not concentrated in specific regions but spread all over the genome, reads matching a reference alignment region likely show a decent amount of variation, making placements exploitable \citep{Linard2020}.

Lastly, the RSs need to at least span the genomic region that the QSs come from.
For a more robust inference of the RT however, it can be advantageous to include a larger region with more phylogenetic signal.
Theoretically, if one wanted to place shotgun sequences from entire genomes, whole-genome RSs would be needed.

As an alternative to manual selection, the Phylogenetic Automatic Reference Tree \citep[PhAT,][]{Czech2018} is a method
that uses reference taxonomic databases 
to select suitable RSs
which represent the diversity of (subsets of) the database.
In cases where taxonomic resolution at the species-level 
does not require expert curation,
the PhAT method can provide a basis for rapid data exploration,
and help to obtain an overview of the data and its intrinsic  diversity.



\paragraph{Reference Alignment Computation}
\label{sec:PhylogeneticPlacement:sub:ReferenceSequencesAlignmentTree:par:AlignmentComp}

Next, for ML-based tree inference and placement, the RSs need to be aligned against each other to obtain the reference alignment (RA).
Typically, this is conducted with \textit{de novo} multiple sequence alignment tools such as
\toolname{T-Coffee} \citep{Notredame2000}, \toolname{MUSCLE} \citep{Edgar2004}, \toolname{MAFFT} \citep{Katoh2002},
and others; see \citep{Kemena2009,Pervez2014,Chatzou2016} for reviews. Recently, \toolname{MUSCLE~v5} introduced an interesting new approach that generates alignment ensembles to capture alignment uncertainty \cite[preprint]{Edgar2021}.
In the ML framework, the QSs also need to be aligned against the RA, see next section.


\paragraph{Tree Inference}
\label{sec:PhylogeneticPlacement:sub:ReferenceSequencesAlignmentTree:par:TreeInference}

Finally, given the RA, a phylogenetic tree of the RSs is inferred, which is henceforth used as the reference tree (RT);
see \citep{Kapli2020} for a general review on this topic.
In theory, any method that yields a fully resolved (bifurcating) tree is applicable,
e.\,g., neighbor joining \cite{Saitou1987},
maximum parsimony \cite{Sankoff1975},
or Bayesian inference \cite{Holder2003,Yang2006}.
In practice however, maximum likelihood (ML) tree inference \citep{Yang2006,Dhar2016} is preferred,
in particular when using ML-based placement,
as otherwise inconsistencies in the assumed models of sequence evolution can affect placement accuracy.
To this end, common software tools include \toolname{IQ-TREE} \cite{Nguyen2015a}, \toolname{FastTree2} \cite{Price2010},
and \toolname{RAxML} \cite{Stamatakis2014,Kozlov2019a};
see \citep{Zhou2018} for a review and evaluation of ML-based tree inference tools.
An open research question in this context is how to incorporate uncertainty in the tree inference (and in the alignment computation) with phylogenetic placement \citep{Huelsenbeck2001,Ronquist2004,Edgar2021}.




\paragraph{Alignment of Query Sequences}
\label{sec:PhylogeneticPlacement:sub:ML:par:Alignment}

For many placement methods, the query sequences need to be aligned against the reference alignment.
In principle, de novo alignment methods can be deployed
to obtain a comprehensive alignment of both the reference and query sequences.
These tools are however not intended for HTS data,
and are not well suited for handling the heterogeneity of phylogenetic placement data,
with (typically) longer, curated, high-quality reference sequences,
and short lower-quality reads (query sequences).


Hence, with the rise of high-throughput sequencing, specialized tools have been developed
that extend a given (reference) alignment without fully recomputing the entire alignment.
In the context of phylogenetic placement, there are two additional advantages that can be exploited
to improve efficiency:
(i)~query sequences only need to be aligned against the reference, but not against each other
(as their phylogenetic relationship is not resolved during placement), and
(ii)~insertions into the reference that result from aligning a QS against the reference 
can be omitted as they do not contain any phylogenetic signal for the placement of the QS.

In the simplest case, only the reference alignment and query sequences are required as input.
For instance, the \codeline{hmmalign} command of \toolname{HMMER} \cite{Eddy1995,Eddy1998}
can align query sequences to the reference alignment using a profile Hidden Markov Model (HMM)
built from the reference alignment.
Note that the option \codeline{-m} has to be set in order to not insert columns of gaps into the reference.
Alternatively, the \toolname{mafft} command \codeline{--addfragments} \citep{Katoh2012}
uses an internally constructed guide tree built from a pairwise distance matrix
of the reference alignment to aid the alignment process;
here, the option \codeline{--keeplength} has to be set to not add columns of gaps to the reference.

Furthermore, the \toolname{PaPaRa} tool \citep{Berger2011a,Berger2012}
can be used that was was specifically developed to target phylogenetic placement.
It takes the RT as additional input,
and uses inferred ancestral sequences at the inner nodes of the tree to improve the alignment process.
Here, the option \codeline{-r} has to be set to not insert columns of gaps into the reference.
Similarly, \toolname{PAGAN} \citep{Loytynoja2012} also utilizes the information in the reference tree,
but it \emph{does} extend the reference alignment with gaps as needed for the query sequence,
causing higher computational effort during placement.

Note that typically, read mapping tools such as \toolname{Bowtie2} \citep{Langmead2012} or
\toolname{BWA} \citep{Li2009,Li2010} are not recommended for phylogenetic placement,
as they expect low-divergent sequences as input, e.\,g., from a single species.

\begin{center}
\begin{table}
\caption{General purpose placement tools. \textmd{
    This table compares the features of the general purpose (i.\,e., not use-case specific) phylogenetic placement tools.
    Columns are as follows.
    Alignment: Does the tool need the QSs to be aligned against the reference alignment?
    Multiple: Does the tool produce multiple placement locations per QS, or just a single (best) one?
    Uncertainty: Is there some measure of uncertainty (such as LWR) assigned to each placement location?
    Branch Length: Does the tool compute the involved branch lengths at each placement location for each QS?
}}
\begin{tabular}{lcccc}
Placement Tool & Alignment & Multiple & Uncertainty & Branch Lengths\\
\hline
\toolname{pplacer}      & yes & yes & yes & yes \\
\toolname{RAxML-EPA}    & yes & yes & yes & yes \\
\toolname{EPA-ng}       & yes & yes & yes & yes \\
\toolname{RAPPAS}       & no  & yes & yes & no  \\
\toolname{APPLES}       & no  & no  & no  & yes \\
\toolname{App-SpaM}     & no  & no  & no  & yes \\
\end{tabular}
\label{tab:PlacementTools}
\end{table}
\end{center}


\subsection{General Purpose Placement Methods}
\label{sec:PhylogeneticPlacement:sub:GeneralPurpose}

Once initial tasks such as reference tree creation and sequence alignment are completed, the actual placement can commence.
There exist several distinct algorithmic approaches for conducting the core part of phylogenetic placement, which we introduce here; see~\tabref{tab:PlacementTools} for an overview.


\paragraph{Maximum Likelihood Placement}
\label{sec:PhylogeneticPlacement:sub:GeneralPurpose:par:ML}

Maximum Likelihood (ML) is a statistically interpretable and robust general inference framework,
and one of the most common approaches for phylogenetic tree inference \citep{Felsenstein2004,Yang2006,Dhar2016}.
It works by searching through the super-exponentially large space of potential tree topologies for a given set of sequences (taxa),
and computing the phylogenetic likelihood of the sequence data of these taxa
being the result of the evolutionary relationships between the taxa as described by each potential tree,
while also computing branch lengths of the tree.
The result of this inference is the tree topology one is able to find using some heuristic search strategy that best (most likely) ``explains'' the underlying sequence data.
Due to the NP-hardness of the tree search problem, the best tree one can find might not be the globally best one. 

To calculate this likelihood, ML methods use statistical models of sequence evolution that describe substitutions between sequences (insertions and deletions are mostly ignored; it is hence also called a substitution model), see \citep{Arenas2015} for a review.
Consequently, the estimated parameters of these models are an inherent property of the resulting phylogenetic tree.
The choice of model parameters also directly informs the specific branch lengths of a tree, interpreting a tree under a different set of model parameters thus may lead to inconsistencies.
Therefore, under the ML framework, we strongly recommend to use the same substitution model and parameters for tree inference and for phylogenetic placement.

Based on the general ML tree inference framework, ML-based phylogenetic placement works in two steps:
First, the QSs are aligned against the RA as described above,
and second, using the resulting comprehensive alignment with both reference and query sequences,
the QSs are placed on the RT using the maximum likelihood method to evaluate possible placement locations \citep{Stark2010,Matsen2010,Berger2011}.





Standard methods used in ML tree inference use search heuristics to explore some possible tree topologies for a given set of sequences. 
Instead, for a given QS, ML-based placement only searches through the branches of the reference tree (RT) as potential placement locations for the QS.
That is, each branch of the RT is evaluated as a placement location, and branch lengths of the involved branches are optimized, following the same approaches as for de novo tree inference.
However, the distal and proximal branch lengths of the placement (see \figref{fig:Placement:sub:TinyTree} for details)
are typically re-scaled, so that their sum is equal to the original branch length in the RT.
Finally, the phylogenetic likelihood of the tree with the QS amended as a temporary extra taxon is calculated.

For each QS and each branch of the RT, this process yields a likelihood score
(which is stored in the \fileformat{jplace} format, see \secref{sec:PhylogeneticPlacement:sub:Concept:par:FileFormat}).
The Likelihood Weight Ratio (LWR) of a placement location is then computed
as the ratio between this likelihood score
and the sum over all likelihood scores for the QS across the entire tree \cite{Strimmer2002,VonMering2007}.
These likelihood scores sum to one across all branches, and hence express the confidence (or probability) of the QS being placed on a given branch.


The first two tools to conduct phylogenetic placement in an ML framework
were the simultaneously published (as preprints) \toolname{pplacer} \citep{Matsen2010} and \toolname{RAxML-EPA} \citep{Berger2011}.
Both build on the same general ML concepts, but employ different strategies 
for improving computational efficiency,
e.\,g., by heuristically limiting the number of evaluated branches (potential placement locations).
Additionally, \toolname{pplacer} offers a Bayesian placement mode.
The more recent \toolname{EPA-ng} \citep{Barbera2018} tool combines features from both \toolname{pplacer} and \toolname{RAxML-EPA},
is substantially faster and more scalable on large numbers of cores, and hence is the recommended tool for ML-based placement.





\paragraph{Ancestral-Reconstruction-Based Placement}
\label{sec:PhylogeneticPlacement:sub:GeneralPurpose:par:AncestralReconstruction}

Recently, multiple methods were introduced that do not rely on aligning query sequences to a reference MSA.
The first such group of methods is based on reconstructing ancestral states at interior nodes of the reference tree, again using an ML framework. 
From these ancestral sequences, $k$-mers are generated and associated with the branches of the reference tree.
Subsequently, phylogenetic placement is performed by comparing the constituent $k$-mers of a QS with the set of $k$-mers indexing the reference tree branches, thereby obviating the need for QS alignment.
This is the general approach used in both \toolname{RAPPAS} \citep{Linard2018} and \toolname{LSHplace} \citep{Brown2012}.

It should be noted that using this procedure, distal and pendant branch lengths of a given RT branch are determined during the association of $k$-mers with RT branches, meaning that all placements on a given branch have the same fixed location.
This means that an additional step to conduct branch length optimization that is not directly offered by \toolname{RAPPAS} or \toolname{LSHplace} may be required to obtain more realistic placement branch lengths. 
\toolname{RAPPAS} however does produce multiple placements per QS and calculates a confidence measure akin to the LWR, yielding a distribution for placing a single QS onto different branches of the tree.


\paragraph{Distance-Based Placement}
\label{sec:PhylogeneticPlacement:sub:GeneralPurpose:par:DistanceBasedPlacement}

Finally, the most recent placement approaches utilize methods from distance-based phylogenetic inference.

For example, \toolname{APPLES} \citep{Balaban2019} is based on the least-squares criterion for tree reconstruction \citep{Felsenstein2004}.
For a given tree, the least-squares method calculates the difference between the pairwise sequence distances and the pairwise patristic distances (i.\,e., the path lengths between two leaves).
A least-squares optimal tree is the tree for which this difference is minimized.
In \toolname{APPLES}, this criterion is used to score possible placement locations of a QS on an existing tree, returning the branch which minimizes the between-distances difference.
A key advantage of the least-squares approach is its ability to efficiently handle reference trees with hundreds of thousands of leaves, which is currently not computationally feasible using ML methods.
Further, the method does not require an alignment of the sequences involved, requiring only a measure of pairwise distance between them.
Note however that as these methods still require a reference tree, computing a reference MSA may still be needed, unless the tree is inferred via distance-based methods as well.
Consequently, even unassembled sequences, such as genome skims \citep{Dodsworth2015}, may be used both as reference and query sequences.
Recently, an updated \toolname{APPLES-2} was published that further improves upon the scalability and accuracy of the tool \citep{Balaban2021}.
Note also that \toolname{APPLES} can take as input, but does not require, aligned sequences.

The most recent alignment-free method is \toolname{App-SpaM} \cite{Blanke2021}.
It utilizes the concept of a spaced-word, which can be understood as a type of $k$-mer for which only some characters have to be identical for two subsequences to be considered as having the same $k$-mer.
This relaxed equality definition is informed by a binary pattern, indicating for each site of a spaced word whether it should be taken into account ($1$) or disregarded ($0$).
Building on this, the tool calculates pairwise distances between a QS and the RSs based on the number of shared spaced-words.
Subsequently, the tool identifies the placement branch of a QS as either the terminal branch of the closest RS, or the branch leading to the parental node of the LCA of the two closest RSs, depending on the strength of the signal of the closest RS.
Notably, \toolname{App-SpaM} is able to provide both distal and pendant branch lengths for the placements it produces, and does so using an estimated phylogenetic distance \cite[the Jukes-Cantor distance,][]{jukes1969evolution}.
Note that both \toolname{APPLES} and \toolname{App-SpaM} only produce a single placement per QS and can therefore not offer statistical measures of placement uncertainty such as the LWR.

Generally, distance-based placement methods produce results with lower accuracy compared to ML-based placement, though this gap appears to be narrowing.
These newer approaches do however expand the scope of placement to sizes of reference trees, and lengths of reference sequences, that are orders of magnitude larger than what is currently possible with ML methods.



\subsection{Application-Specific Placement Methods}
\label{sec:PhylogeneticPlacement:sub:Alternative}

Several additional placement methods exist. We provide a survey of these in this section.
The placement methods covered in this section set themselves apart through their more specific use-cases, however this does not imply that their scope of use is necessarily limited.


\paragraph{Viral Data}
\label{sec:PhylogeneticPlacement:sub:Alternative:par:viral}

A particularly challenging use case for phylogenetic methods is the investigation of viral data, with a highly relevant example coming from the SARS-CoV-2 pandemic.
Due to the dense sampling involved in studying such viral outbreaks, differences between individual taxa in a prospective tree may only be due to a very low number of, or even single, mutations.
Consequently the amount of phylogenetic signal is generally very low, complicating tree reconstruction \citep{Morel2020sars}.
Yet, distinguishing between major viral variants and identifying them precisely from a given clinical sample is crucial for epidemiological studies.
In this context the \toolname{UShER} software was introduced that specifically focuses on phylogenetic placement of SARS-CoV-2 sequences \citep{Turakhia2021}.
In contrast to ML methods, \toolname{UShER} uses a Maximum Parsimony (MP) approach, and does not operate on the full sequence alignment.
This allows the method to focus directly on individual mutations, and consequently only use a fraction of the runtime and memory footprint of conventional ML placement methods.
Note that the accuracy of MP-based phylogenetic methods can suffer when one or more lineages in the tree have experienced rapid evolution that results in long branch lengths.
In such cases MP may incorrectly determine such lineages to be closely related, an effect termed \emph{long branch attraction} \citep{Felsenstein1978,Bergsten2005}.
While this is less of an issue for very closely related sequences such as SARS-CoV-2 or other (but not all) viral data, it may yield the application of such approaches to different types of data more challenging.


\paragraph{Gene Trees}
\label{sec:PhylogeneticPlacement:sub:Alternative:par:Genetrees}

In principle, all placement methods aim to provide the location of a QS on a phylogeny that accurately reflects the underlying pattern of speciation, i.\,e., the \emph{species tree}.
In practice, the reference tree is typically only inferred on a single gene (16S, 18S, ITS, etc.), yielding a \emph{gene tree} which may substantially differ from the species tree, called gene-tree \emph{discordance} \citep{Degnan2009}.
Alternatively, we may have multiple such gene trees that induce a species tree, and subsequently want to perform query placement onto the species tree via placement onto the constituent gene trees \citep{Sunagawa2013a}.
Currently, only two placement methods are able to handle such cases:
\toolname{INSTRAL} and \toolname{DEPP}.
\toolname{INSTRAL} \citep{Rabiee2019} performs placement of QSs for a species tree induced by a set of gene trees.
It does so by first placing into the individual gene trees using existing ML placement methods, then re-inferring the species tree from the extended gene trees.
In contrast to this, \toolname{DEPP} \citep[preprint]{Jiang2021} only considers the problem of discordance between a gene tree and its species tree and attempts to account for this during the placement into the species tree.
The approach is based on a model of gene tree discordance learned from the data using deep neural networks that yields an embedding of given sequences into a euclidean space.
Incidentally, this makes \toolname{DEPP} the first and so-far only phylogenetic placement method to incorporate machine learning.
\toolname{DEPP} then uses the pairwise distances that result from the embedding of both reference and query sequences as input to \toolname{APPLES}, which computes the least-squares placement of the QSs.


\paragraph{Other Use Cases}
\label{sec:PhylogeneticPlacement:sub:Alternative:par:other}

Some further tools make application-specific usage of placement.
The first pertains to the specific case of samples containing sequences from exactly two organisms, and the task of identifying their respective known reference organisms.
The tool \toolname{MISA} was developed with this specific use-case in mind \citep{Balaban2020}.

The second relates to either placing morphological sequences from fossils typically represented by binary characters (presence/absence of a trait) or Ancient DNA (aDNA) sequences.
Placing ancient DNA sequences is generally challenging for analysis because of the high degree of degradation due to the age of the DNA molecules, generally shorter read lengths ranging between 50 and 150 base pairs, and post-mortem deamination \citep{Hofreiter2001}.
The \toolname{pathPhynder} tool aims to solve this use-case \citep[preprint]{Martiniano2020}.
Like \toolname{UShER}, \toolname{pathPhynder} operates on nucleotide variants, focusing on single nucleotide polymorphisms.
Furthermore, phylogenetic placement has been used for placement of fossils \citep{Berger2010a,Bomfleur2015} using morphological data. This approach uses the maximum likelihood framework to use the signal from mixed morphological (binary) and molecular partitions in the underlying MSA.

Lastly, phylogenetic placement has also been proposed as a way to perform OTU clustering.
The \toolname{HmmUFOtu} \citep{Zheng2018} tool implements this specific use-case, along with automated taxonomic assignment (see also \secref{sec:PhylogeneticPlacement:sub:TaxonomicClassification}).
A unique characteristic in comparison to other placement tools is that \toolname{HmmUFOtu} also performs QS alignment and uses this information to pre-select promising placement locations.


\subsection{Workflows based on Phylogenetic Placement}
\label{sec:PhylogeneticPlacement:sub:RelatedTools}

Over the last decade, several pipelines have been published that use phylogenetic placement tools as their core method, building on it and using its result in various ways.


\paragraph{Automated Analysis Pipelines}
\label{sec:PhylogeneticPlacement:sub:RelatedTools:par:Pipelines}

One class of placement pipelines focus on simplifying the overall use of placement methods, typically providing the user with the option to use a pre-computed reference tree, obviating the need for manual selection of reference taxa \citep{Stark2010,Carbone2016,Carbone2019,Douglas2018,Douglas2020a,Erazo2021,Sempr2021}.
A number of these pipelines also automate the generation of key metrics and downstream analysis steps.
Among these pipelines, of particular note is \toolname{PICRUSt2} \citep{Douglas2018, Douglas2020a}, which stands out for accounting for 16S copy number correction, and providing the user with a prediction of the functional content of a sample.
Similarly, \toolname{paprica} \citep{Erazo2021} is a pipeline that computes metabolic pathway predictions for bacterial metagenomic sample data.



\paragraph{Divide-and-Conquer Placement}
\label{sec:PhylogeneticPlacement:sub:RelatedTools:par:DivideandConquer}

A further key challenge for existing phylogenetic placement tools is scalability with regards to the size of the reference tree.
While more recent methods have shown significant improvements in both the memory footprint and execution time required when placing QSs on reference trees on the order of $10^5$ reference taxa (see \secref{sec:PhylogeneticPlacement:sub:GeneralPurpose:par:DistanceBasedPlacement}), such input sizes remain extremely challenging for ML-based placement methods.
A number of workflows have been proposed to scale existing placement methods for this use-case by splitting up the reference tree into smaller subtrees on which phylogenetic placement is then performed, creating a divide-and-conquer approach to phylogenetic placement \citep{Mirarab2012,Czech2018,Czech2020,Koning2021,Wedell2021}.
These approaches vary primarily in how they select subtrees.
\toolname{SEPP} \citep{Mirarab2012} and \toolname{pplacerDC} \citep{Koning2021} generate a subtree based on the topology of the reference tree.
\toolname{SEPP} is a general boosting technique in particular for highly diverse reference trees \citep{Mirarab2012,Liu2012}.
Further, a multi-level placement approach exists \citep{Czech2018,Czech2020}, which first places onto a broad RT, and then extracts QSs in pre-selected clades of that RT to place them again onto clade-specific high-resolution RTs.
Finally, \toolname{pplacer-XR} \citep{Wedell2021} selects a set of neighboring reference branches based on similarity to each query sequence, out of which it creates a subtree.
Note that in this case, when decomposing the reference tree differently for every query sequence, scalability with regards to the number of query sequences is severely reduced.

A central promise of placement on very large trees is to simplify the curation and engineering tasks involved in creating a reference tree, as here a typical challenge is to decide which taxa to include in the tree.
If placement can instead be performed on a tree encompassing an entire database, the curation challenge is circumvented. 
However, as another common issue with reference tree generation is the inclusion of overly similar reference sequences resulting in unclear or fuzzy placement signal, divide-and-conquer placement approaches may not be sufficient on their own.

\paragraph{Evaluation of Placement Tools}
\label{sec:PhylogeneticPlacement:sub:RelatedTools:par:Evaluation}

Lastly, \toolname{PEWO} is an extensible testing framework specifically aimed at benchmarking and comparing different phylogenetic placement softwares \citep{Linard2020}.
It includes a wide range of datasets and thus provides an important resource for identifying which placement tool is best suited for specific use-cases by evaluating the accuracy of existing tools, given some dataset.
\toolname{PEWO} does so using a pruning-based evaluation procedure, where a subset of leaves is removed from a reference tree.
This subset of sequences is subsequently used as input QSs for placement.
The accuracy of a placement is calculated as the number of nodes between the best placement location, and the original location of the QS on the reference tree (called the node distance).
This basic approach is used for evaluation in most publications that introduce new placement approaches.
Note that the node distance measures two sources of error: error introduced by the placement algorithm, and error introduced by the pruning of the reference tree.
In contrast to this, the ``delta error'' used in the evaluation of \toolname{APPLES} measures the additional error introduced through placement, in addition to the error introduced by the process of altering the reference tree through pruning \citep{Balaban2019}.
This new metric is however not yet included in the \toolname{PEWO} workflow.
Nevertheless, the usefulness of a comprehensive and standardized testing framework cannot be emphasized enough, as it substantially facilitates further advancement and standardization in the field and the development of novel methods.



\section{Visualization and Analysis}
\label{sec:PlacementAnalysis}

As mentioned before, there are two ways to conceptualize phylogenetic placement:
(i) as an assignment (or mapping) of individual sequences to the branches of a phylogeny,
usually taking the ($n$-)most likely placement location(s) of each sequence, or
(ii) as the distribution of all sequences of a sample across the tree,
taking their respective abundances and placement probabilities into account.
The former is similar to taxonomic assignment, but with full phylogenetic resolution instead of resolution at the taxonomic levels only,
while the latter focuses on, e.\,g., species communities and their diversity as a whole. In the following we provide an overview of analysis methods that make use of such data.


\paragraph{Abundances and Multiplicities}
\label{sec:PlacementAnalysis:par:Abundances}

In both interpretations, an important consideration is whether to take sequence abundances into account.
When working with strictly identical sequences, or sequences resulting from some (OTU) clustering, the number of occurrences of each sequence or size of each cluster
can be used as additional information for interpreting, e.\,g., community structure.
On the one hand, including their abundances with the placement of each sequence yields information on how prevalent the species of these sequences are;
for example, this can provide insight into the key (most abundant) species in environmental samples.
On the other hand, dropping abundances and instead considering each sequence once (as a singleton) is more useful for estimating total diversity and taxonomic composition.
For example, this way the number of \emph{distinct} sequences can be regarded as a proxy for the number of species that are present in a sample.
Whether to include abundances should hence be decided depending on the type of analysis conducted.

In the \fileformat{jplace} format, these abundances can be stored
as the so-called ``multiplicity'' of each placement \citep{Matsen2012}, in the \verb|"nm"| data field.
Unfortunately, the \fileformat{fasta} \citep{Pearson1988} and \fileformat{phylip} \citep{Felsenstein1981} formats
used as input to placement do not natively support abundance annotations,
and current placement tools often do not handle them automatically, meaning that the information can be lost.
However, the \toolname{chunkify} workflow \citep{Czech2018,Czech2020} mentioned in \secref{sec:PhylogeneticPlacement:sub:TypesOfSequences:par:Clustering} takes abundances into account and annotates them as multiplicities in the resulting \fileformat{jplace} file.
Furthermore, \toolname{gappa} \citep{Czech2020} offers a command to edit the multiplicities as needed, for example setting them post-hoc to the initial sequence abundance determination.


\paragraph{Visualization}
\label{sec:PlacementAnalysis:par:Visualization}

Prior to more in-depth analyses, a first step in most workflows is a visualization of the immediate results.
Following the two interpretations of phylogenetic placement (and hence, depending on the research question at hand), there are several ways to visualize placement results.

First, individual placements can be shown as actual branches attached to the RT,
e.\,g., \figref{fig:Placement:sub:PlacementLocation}.
Typically, only the most likely placement location per sequence is used for this,
in order to avoid cluttering of the tree; this hence omits the information about uncertainty.
This can be conducted by generating trees from placement results, e.\,g., in \fileformat{newick} format.
Tools to this end are \toolname{gappa} \citep{Czech2020} and \toolname{guppy}, which is part of \toolname{pplacer} \citep{Matsen2010}.
This can subsequently be visualized via standard tree viewing tools \citep[for a review, see][]{Czech2017}.
Note however that such a visualizations can quickly become overloaded when the number of QSs becomes large.

Second, the LWR distribution of a single sequence can be visualized, to depict the uncertainty in placement across the tree, for example with \toolname{ggtree} \citep{Yu2017} and \toolname{iTOL} \citep{Letunic2016,Letunic2019}.

Third, the distribution of \emph{all} sequences can be visualized directly on the reference tree, for example as shown in \figref{fig:Placement:sub:Distribution},
taking their per-branch probabilities (and potentially their multiplicities/abundances) into account.
This gives an overview of all placements, and can for example reveal important clades that received a high fraction of placements,
or indicate whether placements are concentrated in a specific region of the tree.
These visualizations can directly be generated by \toolname{gappa} \citep{Czech2020} and \toolname{iTOL} \citep{Letunic2016,Letunic2019};
furthermore, \toolname{guppy}, can produce tree visualizations in the \fileformat{phyloXML} format \citep{Han2009}, which can subsequently be displayed by tree viewer tools such as \toolname{Archaeopteryx} \citep{Han2009}.



\begin{figure*}[!tb]
    \centering
    \includegraphics[width=\linewidth]{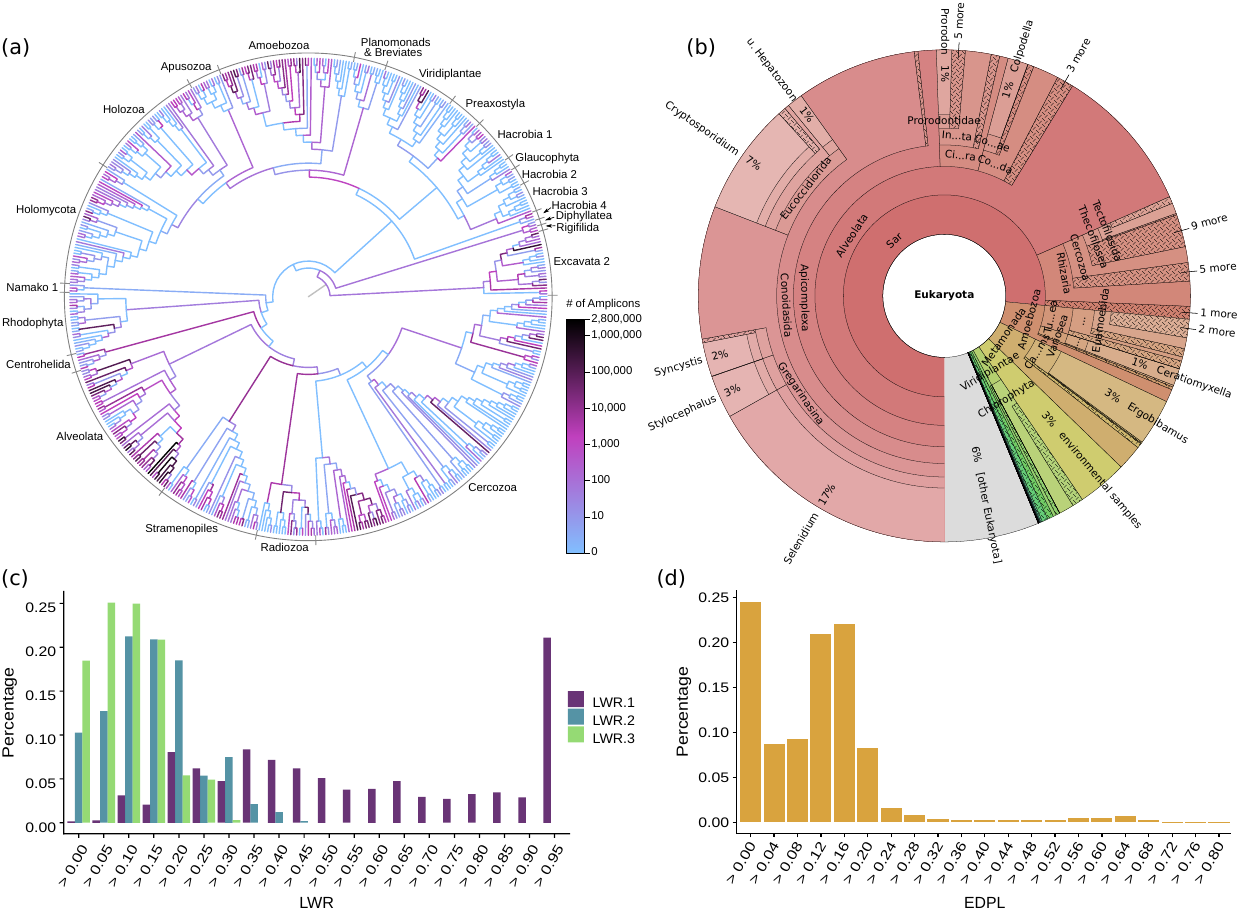}
    \begin{subfigure}{0pt}
        \phantomcaption
        \label{fig:Examine:sub:MassTree}
    \end{subfigure}
    \begin{subfigure}{0pt}
        \phantomcaption
        \label{fig:Examine:sub:Krona}
    \end{subfigure}
    \begin{subfigure}{0pt}
        \phantomcaption
        \label{fig:Examine:sub:LWRHistogram}
    \end{subfigure}
    \begin{subfigure}{0pt}
        \phantomcaption
        \label{fig:Examine:sub:EDPL}
    \end{subfigure}
    \vspace*{-1em}
    \caption{
        \textbf{Examination of phylogenetic placement data.}
        Here, we show some techniques for visually inspecting placement data,
        using an exemplary dataset consisting of 154 soil samples from neotropical rain forests
        placed on an eukaryotic reference tree \citep{Mahe2017}.
        \hspace{1em}\subref{fig:Examine:sub:MassTree}
        Heat tree showing the distribution of millions of amplicon reads on the reference tree
        by summing over the per-branch Likelihood Weight Ratios (LWRs) of all reads.
        The high abundance of placed reads in the \entityname{Alveolata} clade (dark branches in the lower left)
        visualizes a main finding of the dataset in form of an over-abundance of reads from that clade,
        shown in the phylogenetic context of the reference tree. Figure adapted from \citep{Mahe2017}.
        \hspace{1em}\subref{fig:Examine:sub:Krona}
        Taxonomic assignment of all reads based on the PR\textsuperscript{2} \citep{Guillou2012,Benson2009} taxonomy.
        The taxonomy of the reference sequences was used to label each branch of the reference tree
        by its highest non-conflicting taxonomic path.
        Then, for each read, the LWRs of its placement locations were accumulated for the branches,
        creating an overview of taxonomic abundances taking placement confidences into account.
        The result across all reads is shown here as a Krona plot \citep{Ondov2011}.
        \hspace{1em}\subref{fig:Examine:sub:LWRHistogram}
        Histogram of the LWRs of the first three most likely placement locations of each read,
        showing how many of the reads have their first, second, and third most likely placement
        at each (binned) LWR value.
        For example, the highest bin of LWR.1 on the right hand side indicates that ~20\% of the reads have a first
        (most likely) placement position at or above an LWR of 0.95.
        That is, these placements have a high LWR and are hence placed with high certainty onto their respective branches.
        Note that the second most likely placement (LWR.2) can never have an LWR exceeding 1/2
        (otherwise, it would be the most likely), the third most likely (LWR.3) not more than 1/3
        (otherwise, it would be the second most likely), and so forth.
        \hspace{1em}\subref{fig:Examine:sub:EDPL}
        Histogram of the Expected Distance between Placement Locations (EDPL),
        which are computed as the distances (in terms of ML branch path length) between placement locations of a query sequence,
        weighted by the respective LWR of each location.
        The EDPL measures how far the placements of a sequence are spread across the branches of the reference tree,
        and hence how certain the placement in a ``neighborhood'' of the tree is.
        Here, most reads have an EDPL below 0.24 branch length units (mean expected number of substitutions per site).
        This indicates that the reads have most of their likely placements close to one another, within two branches on average,
        given that the used reference tree has an average branch length of about 0.12.
    }
    \label{fig:Examine}
\end{figure*}


\subsection{Placement Quality and Uncertainty Quantification}
\label{sec:PlacementAnalysis:sub:UncertaintyQuantification}

An important post-analysis aspect is quality control,
both in order to assess the suitability of the RT for the given placed sequences
(to, e.\,g., test for missing reference sequences), and in order to assess the placed sequences themselves.
Assuming a `perfect' reference tree that exactly represents the diversity of the query sequences,
the theoretical expectation is that each sequence gets placed onto a leaf of the tree with an LWR close to 1.
Ignoring sequencing errors and other technical issues,
deviations from this expectation can be due to several issues. 

To this end, plotting the histograms or the distribution of the confidences (LWRs) across all placements can be useful,
\figref{fig:Examine:sub:LWRHistogram}.
A more involved metric is the so-called Expected Distance between Placement Locations \citep[EDPL,][]{Matsen2010}, 
which for a given sequence represents the uncertainty-weighted average distance between all placement locations of that sequence,
or in other words, the sum of distances between locations, weighted by their respective probability,
see \figref{fig:Examine:sub:EDPL}. 
The EDPL is a measure of how far the likely placement locations of a sequence are spread out across the tree.
It hence can distinguish between local and global uncertainty of the placements,
that is, between cases where nearby edges constitute equally good placement locations
versus cases where the sequence does not have a clear placement position in the tree \citep{Matsen2010}.
These metrics can be explored with \toolname{gappa} \citep{Czech2020} and \toolname{guppy} \citep{Matsen2010};
see their respective manuals for the available commands.

Examining the distribution of placement statistics,
\figref{fig:Examine:sub:LWRHistogram}-\subref{fig:Examine:sub:EDPL},
or even the values of individual sequences,
can help to identify the causes of problematic placements:
(i)~Sequences that are spread out across a clade with a flat placement distribution might indicate that too many closely related sequences, such as strains, are included in the RT;
the EDPL can be used to quantify this.
The query sequence is then likely another variant belonging to this subtree.
(ii)~Placements towards inner branches of the RT might hint a hard to place query sequence,
or at a lack of reference sequence diversity.
This occurs if the (putative) ancestor represented by an inner node of the tree
is more closely related to the QSs than the extant representatives included in the RT.
This can either be the result of missing taxa in the RT,
or even because the diversity of the clade is not fully known yet (also known as incomplete taxon sampling), in which case the QS might have originated from a previously undescribed species.
(iii)~Sequences placed in two distinct clades might indicate technical errors such as the presence of chimeric sequences \cite{Haas2011}.
(iv)~Sequences with elevated placement probability in multiple clades (e.\,g., placements in more than two subtrees) usually result from more severe issues, such as a total lack of suitable reference sequences for the QS, or a severe misalignment of the QS to the reference.
This can for instance occur if metagenomic shotgun data has not been properly filtered,
such that the genome region that the QS originated from is not included in the underlying MSA.
(v)~Lastly, long pendant lengths can also occur if a QS does not fit anywhere in the RT,
in particular when the RT contains outgroups,
which can cause long branch attraction for placed sequences \citep{Bergsten2005}.

Quantifying these uncertainties in a meaningful and interpretable way, and distinguishing between their causes,
are open research questions.
Approaches such as considering the EDPL, flatness of the LWR distribution,
pendant lengths relative to the surrounding branch lengths of the RT, might help here,
but more work is needed in order to distinguish actual issues
from the identification of a new species based on their placement.





\subsection{Taxonomic Classification and Functional Analysis}
\label{sec:PhylogeneticPlacement:sub:TaxonomicClassification}

By understanding the taxonomic composition of an environment,
questions about its species diversity and richness can be answered.
Typical metagenomic data analyses hence often include a taxonomic classification of reads
with respect to a database of known sequences \cite{Breitwieser2019},
for example by aggregating relative abundances per taxonomic group.
In addition, such a classification based on known data enables to analyze which pathways and functions are present in a sample, and hence to gain insight into the metabolic capabilities of a microbial community.


\paragraph{Preexisting Tools}
\label{sec:PhylogeneticPlacement:sub:TaxonomicClassification:par:PreexistingTools}

Many tools exist to these ends:
\toolname{BLAST} \citep{Altschul1990} and other similarity-based methods were among the early methods,
but depend on the threshold settings for various parameters \cite{Shah2018},
only provide meaningful results if the reference database contains sequences closely related to the queries \cite{Mahe2017},
and the closest hit does often not represent the most closely related species \citep{Koski2001,Clemente2011}.
Thus, the advantages of leveraging the power of phylogenetics for taxonomic assignment have long been recognized  \cite{Delsuc2020}.
The classification can be based on \emph{de novo} construction of a phylogeny \cite{Krause2008,Schreiber2010},
which as mentioned is computationally expensive, and tree topologies might change between samples, yielding downstream analyses and independent comparisons between studies challenging \citep{Boyd2018}.
Alternatively, dedicated pipelines for 16S metabarcoding data
such as \toolname{QIIME} \citep{Caporaso2010a,Bolyen2019} and \toolname{mothur} \citep{Schloss2009}
are routinely used to conduct taxonomic assignment
based on sequence databases and established phylogenies as well as taxonomies;
see \secref{sec:PhylogeneticPlacement:sub:ReferenceSequencesAlignmentTree:par:SequenceSelection}
for a list of common databases, and see \citep{Lopez-Garcia2018,Prodan2020} for comparisons of such pipelines.
Other tools for taxonomic assignment and profiling are available, for example based on $k$-mers,
which often use a fixed taxonomy such as the NCBI taxonomy \citep{Sayers2009,Benson2009} to propose an evolutionary context for query sequences.
They hence use a taxonomic tree without branch lengths, which can be an advantage when a fully resolved phylogeny is not available.
Tools to this end are for example \toolname{MEGAN} \citep{Huson2007a}, \toolname{Kraken2} \citep{Wood2014,Wood2019}, and \toolname{Kaiju} \citep{Menzel2016a}, see \citep{Sczyrba2017,Bremges2018,Meyer2019,Ye2019} for benchmarks and comparisons.
However, these approaches are based on sequence similarity and related approaches, and can therefore be incongruent with the true underlying phylogenetic relationships of the sequences under comparison \citep{Smith2017}.


\paragraph{Placement-Based Approaches}
\label{sec:PhylogeneticPlacement:sub:TaxonomicClassification:par:PlacementBasedApproaches}

Phylogenetic placement can be employed to perform an accurate assignment of QSs to taxonomic labels \citep{Czech2018}, with potentially higher resolution than methods based on manually curated taxonomies \cite{Darling2014,Rajter2021}.
This approach leverages models of sequence evolution \cite{Darling2014},
and is hence more accurate than similarity-based methods \citep{VonMering2007}.
A further advantage over the above pipelines is the ability to use custom reference trees,
thus providing 
a better context for interpreting the data under study.
Incongruencies between the taxonomy and the phylogeny can however hinder the assignment, if they are not resolved \citep{Matsen2012a}.
Furthermore, it is important to note that placement-based methods only work when the query sequences are homologous to the available reference data, hence currently limiting the approach to, e.\,g., short genomes, metabarcoding or filtered metagenomic data.

A simple approach for taxonomic annotation based on placements is to label each branch of the RT by the most descriptive taxonomic path of its descendants, and to assign each QS to these labels based on its placement locations, potentially weighted by LWRs \citep{Kozlov2016,Czech2018}.
This is implemented in \toolname{gappa} \citep{Czech2020}, see \figref{fig:Examine:sub:Krona} for an example;
a similar visualization of the taxonomic assignment of placements can be conducted with \toolname{BoSSA} \citep{bossa}.

More involved and specialized approaches have also been suggested.
\toolname{PhyloSift} \cite{Darling2014} is a workflow that employs placement for taxonomic classification,
using a database of gene families that are particularly well suited for metagenomics.
The workflow further includes \emph{Edge PCA} (introduced in \secref{sec:PlacementAnalysis:sub:SampleAnalysis:par:Similarity}) to assess community structure across samples, and offers Bayesian hypothesis testing for the presence of phylogenetic lineages.
The gene-centric taxonomic profiling tool \toolname{metAnnotate} \citep{Petrenko2015} uses a similar approach to identify organisms within a metagenomic sample that perform a function of interest.
To this end, it searches shotgun sequences against the NCBI database \citep{Sayers2009,Benson2009} first, and then employs placement to classify the reads with respect to genes and pathways of interest.
\toolname{GraftM} \citep{Boyd2018} is a tool for phylogenetic classification of genes of interest in large metagenomic datasets.
Its primary application is to characterize sample composition using taxonomic marker genes, which can also target specific populations or functions.
The abundance profiling methods \toolname{TIPP} \citep{Nguyen2014} and \toolname{TIPP2} \citep{Shah2021} also use marker genes, and employ the \toolname{SEPP} \citep{Mirarab2012,Liu2012} boosting technique for phylogenetic placement with highly diverse reference trees, which increases classification accuracy when under-represented (novel) genomes are present in the dataset.
The more recently introduced \toolname{TreeSAPP} tool \citep{Morgan-Lang2020} uses a similar underlying framework, but improves functional and taxonomic annotation by regressing on the evolutionary distances (branch lengths) of the placed sequences, thereby increasing accuracy and reducing false discovery.
Lastly, \toolname{PhyloMagnet} \citep{Schon2019} is a workflow for gene-centric metagenome assembly (MAGs) that can determine the presence of taxa and pathways of interest in large short-read datasets.
It allows to explore and pre-screen microbial datasets, in order to select good candidate sets for metagenomic assembly.




\subsection{Diversity Estimates}
\label{sec:PhylogeneticPlacement:sub:DiversityEstimates}

A goal that is intrinsically connected to taxonomic assignment in studies that involve metagenomic and metabarcode sequencing is to quantify the diversity within a sample (called $\alpha$-diversity) and the diversity between samples (called $\beta$-diversity).
A plethora of methods exists to quantify the diversity of a set of sequences \citep[for an excellent review, see][]{Tucker2017guide}.
Here, we focus on those approaches that specifically work in conjunction with phylogenetic placement.

Among the $\alpha$-diversity metrics, Faith's Phylogenetic Diversity (PD) stands out, both for its widespread use in the literature and its direct use of phylogenetic information \citep{Faith1992}.
More recently, a parameterized generalization of the PD was introduced that is able to interpolate between the classical PD and its abundance weighted formulation \cite{McCoy2013}.
Notably, this Balance Weighted Phylogenetic Diversity (BWPD) has been implemented to work directly with the results of phylogenetic placement, using the \toolname{guppy} \texttt{fpd} command \citep{Matsen2010,Darling2014}.

To our knowledge, the only other method that computes a measure of $\alpha$-diversity directly from phylogenetic placement results is \toolname{SCRAPP} \citep{Barbera2020a}, which also deploys species delimitation methods \citep{Zhang2013,Kapli2017}.
In this method, the connection of phylogenetics to diversity is through the concept of a molecular species \citep{Agapow2004}, and quantifying how many such species are contained within a given sample.
To facilitate this, \toolname{SCRAPP} resolves the between-QS phylogenetic relationships, resulting in per-reference-branch trees of those QSs that had their most likely placement on that specific branch.
Thus, a byproduct of applying this method is a set of phylogenetic trees of the query sequences.

When the goal is to compute a $\beta$-diversity measure, a common choice for non-placement based approaches is the so-called Unifrac distance \cite{Lozupone2005,Lozupone2007a}, which quantifies the relatedness of two communities that are represented by leaves of a shared phylogenetic tree.
Interestingly, the weighted version of the Unifrac distance has been shown to be equivalent to the KR-distance \citep{Evans2012}, see \secref{sec:PlacementAnalysis:sub:SampleAnalysis:par:Similarity}.
As the Unifrac distance is widely used and well understood, this makes the KR-distance a safe choice for calculating between-sample distances, and thus a measure of $\beta$-diversity based on phylogenetic placement results.


\begin{figure*}[!tb]
    \centering
    \includegraphics[width=\linewidth]{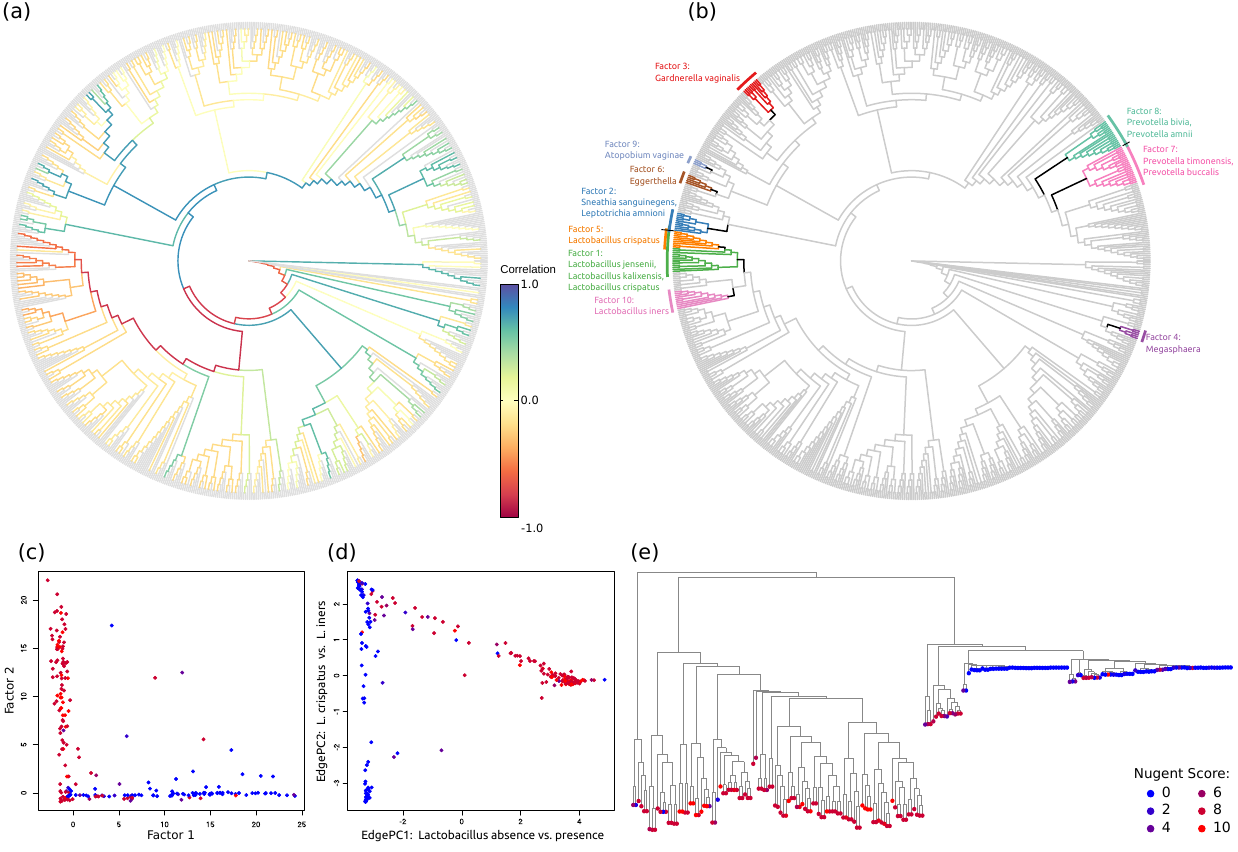}
    \begin{subfigure}{0pt}
        \phantomcaption
        \label{fig:Analyses:sub:Correlation}
    \end{subfigure}
    \begin{subfigure}{0pt}
        \phantomcaption
        \label{fig:Analyses:sub:PlacementFactorizationTree}
    \end{subfigure}
    \begin{subfigure}{0pt}
        \phantomcaption
        \label{fig:Analyses:sub:PlacementFactorizationOrdination}
    \end{subfigure}
    \begin{subfigure}{0pt}
        \phantomcaption
        \label{fig:Analyses:sub:EdgePCA}
    \end{subfigure}
    \begin{subfigure}{0pt}
        \phantomcaption
        \label{fig:Analyses:sub:SquashClustering}
    \end{subfigure}
    \vspace*{-1em}
    \caption{
        \textbf{Analyses of phylogenetic placement data.}
        Here, we show several analysis techniques for placement data,
        which relate multiple samples to each other (e.\,g., from different locations or points in time)
        that have been placed on the same underlying reference tree.
        The example dataset contains 220 vaginal samples of human patients with and without
        Bacterial Vaginosis (BV), a condition caused by an abnormal vaginal microbiome \citep{Srinivasan2012},
        placed on a bacterial tree.
        The ``Nugent'' score is an external clinical indicator of the disease \citep{Nugent1991},
        which is shown in \subref{fig:Analyses:sub:PlacementFactorizationOrdination}--%
        \subref{fig:Analyses:sub:SquashClustering} as blue (healthy, low score) vs red colors (severe disease, high score).
        In healthy patients, two types of \entityname{Lactobacilli} dominate the microbiome,
        while in diseased patients, a diverse mixture of other bacteria take over.
        All figures are adapted from \citep{Czech2019a},
        for details see \citep{Czech2019a,Matsen2011a,Srinivasan2012,Czech2020a}.
        \hspace{1em}\subref{fig:Analyses:sub:Correlation}~%
        \emph{Edge Correlation} between read abundances in clades of the reference tree
        (measured via the \emph{imbalance} transformation) and the per-sample Nugent score.
        This visualization method identifies taxa whose abundances exhibit a relationship with environmental factors.
        Here, the red path towards the left identifies the \entityname{Lactobacillus} clade,
        that exhibits a strong anti-correlation with the Nugent score
        (healthy patients with a low score have high abundances in this clade),
        while blue and green paths show a multitude of clades that correlate with the score
        (diseased patients with a high score and high abundances in these diverse clades).
%
        \hspace{1em}\subref{fig:Analyses:sub:PlacementFactorizationTree}~%
        \emph{Placement-Factorization} discretely identifies these clades
        by splitting up the tree into a number of ``factors'':
        Black edges (with colorized clades below them) indicate the first ten factors (groups of taxa, some of them nested)
        whose differential placement abundances between samples exhibit a strong relationship with the Nugent score.
        That is, a factor is a clade in which abundances co-vary with metadata (e.\,g., the Nugent score).
        Here, these factors are again the \entityname{Lactobacillus} clade and a multitude of other clades
        that are also highlighted in \subref{fig:Analyses:sub:Correlation} by colored paths.
%
        \hspace{1em}\subref{fig:Analyses:sub:PlacementFactorizationOrdination}~%
        \emph{Placement-Factorization} can also ordinate samples, by plotting the \emph{balances}
        (i.\,e., the abundance contrasts) across the edges identified by factors.
        Here, the first two factors of \subref{fig:Analyses:sub:PlacementFactorizationTree}
        are shown (each dot represents one sample, colored by its Nugent score),
        which split healthy and diseased patients.
        \hspace{1em}\subref{fig:Analyses:sub:EdgePCA}~%
        \emph{Edge Principal Components Analysis} (EdgePCA) is another ordination method,
        using PCA on the edge \emph{imbalances}.
        Here, the first two PC axes are shown,
        which separate healthy from diseased patients (\entityname{Lactobacillus} presence vs absence) on the first axis,
        and further distinguish the healthy patients based on the two types of \entityname{Lactobacilli} on the second axis.
        These interpretations of the axes are derived from visualizing the PCA directly on the reference tree, which is another way to show Edge PCA results, see \citep{Matsen2011a,Czech2020a}.
%
        \hspace{1em}\subref{fig:Analyses:sub:SquashClustering}~%
        \emph{Squash Clustering} is a hierarchical clustering method, here showing the clustering tree of the samples (not a phylogeny).
        Tip nodes (leaves) correspond to samples (individual patients), again colorized by their Nugent score,
        with samples clustered based on similarity of their placement distribution,
        and vertical distances showing this similarity, measured as the phylogenetic Kantorovich-Rubinstein (KR) distance between samples.
        Patients with a similar health status are close to each other, in particular the healthy (blue) ones.
    }
    \label{fig:Analyses}
\end{figure*}


\subsection{Placement Distribution}
\label{sec:PlacementAnalysis:sub:PlacementDistribution}

Depending on the research question at hand, and for larger numbers of QSs,
it is often more convenient and easier to interpret to look at the overall placement distribution
instead of individually placed sequences.
This distribution,
as shown in \figref{fig:Placement:sub:Distribution} and \figref{fig:Examine:sub:MassTree},
summarizes an entire sample (or even multiple samples) by adding up the per-branch probabilities (i.\,e., LWRs) of each placement location of all sequences in the sample(s),
ignoring all branch lengths (distal, proximal, and pendant) of the placements.
In this context, the accumulated per-branch probabilities are also called the \emph{edge mass} of a given branch.
This terminology is derived from viewing the reference tree as a graph consisting of nodes and edges,
and viewing the placements as a mass distribution on that graph.
This focuses more on the mathematical aspects of the data,
and provides a useful framework for the analysis methods described below.


\paragraph{Normalization of Absolute Abundances}
\label{sec:PlacementAnalysis:sub:PlacementDistribution:par:NormalizationCompositionalData}

High-throughput meta\-genomic sequence data are inherently compositional \citep{Li2015a,Gloor2017,Quinn2018},
meaning that the total number of reads from HTS (absolute abundances) are mostly a function of available biological material and the specifics of the sequencing process.
In other words, the total number of sequences per sample (often also called library size) is insignificant when comparing samples.
This implies that sequence abundances are not comparable across samples, and that they can only be interpreted as proportions relative to each another \citep{Silverman2017,Calle2019}.
However, the PCR amplification process is known to introduce biases \citep{Logares2014},
potentially skewing these proportions.
For example, the relative abundances of the final amplicons
do not necessarily reflect the original ratio of the input gene regions \citep{Kanagawa2003,Li2015a}; 
this can be problematic in comparative studies.
If these characteristics are not considered in analyses of the data \citep{Weiss2017},
spurious statistical results can occur \citep{Aitchison1986,Jackson1997,Tsilimigras2016,Gloor2016a};
see \citep{Czech2020a} for further details.
For this reason, the estimation of indices such as the species richness is often implemented
via so-called \emph{rarefaction} and rarefaction curves \cite{Gotelli2001}, which might however ignore a potentially large amount of the available valid data \cite{McMurdie2014}.

Phylogenetic placement of such data hence also needs to take this into account.
The total edge masses (e.\,g., computed as the sum over all LWRs of a sample) are not informative,
and merely reflect the total number of placed sequences. 
A simple strategy, upon which several of the analysis methods introduced below are based,
is the normalization of the masses by dividing them by their total sum,
effectively turning absolute abundances into relative abundances.
This also eliminates the need for rarefaction, as low-abundance sequences only contribute marginally to the data.
However, using this approach can still induce compositional artifacts in the data,
as the per-branch probabilities (and hence the edge masses per sequence) have to sum to one for all branches of the tree.
In other words, it is conceptually not possible to change the relative edge mass on a branch
without also affecting edges masses on other branches.


\paragraph{Transformations of Compositional Data}
\label{sec:PlacementAnalysis:sub:PlacementDistribution:par:Transformations}

A statistically advantageous way to circumvent these effects, and resulting misinterpretations of compositional placement data, is to transform the data from per-branch values to per-clade values.
This way, individual placement masses in the nearby branches of a clade are transformed into a single value for the entire clade, which expresses a measure of difference (called contrast) of the placement masses within the clade versus the masses in the remainder of the tree.
This makes such transformations robust against placement uncertainty in a clade (e.\,g., due to similar reference sequences), implicitly captures the tree topology, and solves the issues of compositional data.
From a technical point of view, this transforms the data from a compositional space into an Euclidean coordinate system \cite{Egozcue2005}, where the individual dimensions of a data point are unconstrained and independent of each other.
This can be achieved by utilizing the reference tree,
whose branches imply bi-partitions of the two clades that are split by each branch \cite{Pawlowsky-Glahn2015,Silverman2017}.
Instead of working with the per-branch placement masses,
the accumulated masses on each side of a branch are contrasted against each other.
This yields a view of the data that summarizes all placements in the clades implied by each branch.
These transformations are, for example, achieved via two methods that in the existing literature
have unfortunately confusingly similar names: imbalances and balances \citep{Czech2020a}.

The edge \emph{imbalance} \citep{Matsen2011a} is computed on the normalized edge masses of a sample:
For each edge, the sum over all masses in the two clades defined by that edge are computed;
their difference is then called the \emph{imbalance} of the edge.
The edge \emph{balance} \citep{Silverman2017,Czech2019a} is computationally similar, but instead of a difference of sums,
it is computed as the (isometric) log-ratio of the geometric means of the masses in each clade;
the resulting coordinates are called \emph{balances} \cite{Egozcue2003,Egozcue2005,Quinn2018}.
Both transformations yield a contrast value for each (inner) branch of the tree,
which can then, for example, be used to compare different samples to each other, see \secref{sec:PlacementAnalysis:sub:SampleAnalysis}.
They differ in the details of their statistical properties,
but more work is needed to examine the effects of this on placement analyses \citep{Czech2020a};
in practice, both can be  (and are) used to avoid compositional artifacts.



\subsection{Analysis of Multiple Samples}
\label{sec:PlacementAnalysis:sub:SampleAnalysis}

In typical metagenomic and metabarcoding studies, more than one sample is sequenced,
e.\,g., from different locations or points in time of an environment.
Furthermore, often per-sample metadata is collected as well,
such as the pH-value of the soil or the temperature of the water where a sample was collected. These data allow
to infer connections between the species community composition of the samples and environmental features.
Given a set of samples (and potentially, metadata variables),
an important goal is to understand the community structure \citep{Tyson2004}.
To this end, fundamental tasks include measuring their similarity (a \emph{distance} between samples),
clustering samples that are similar to each other according to that distance measure,
and relating the samples to their environmental variables.
To this end, the methods introduced in this section utilize phylogenetic placement,
and assume that the sequences from all samples have been placed onto the same underlying reference tree;
they are implemented in \toolname{gappa} \citep{Czech2020} and partially in \toolname{guppy} \citep{Matsen2010}.


\paragraph{Similarity between Samples}
\label{sec:PlacementAnalysis:sub:SampleAnalysis:par:Similarity}


A simple first data exploration method consists in computing the \emph{Edge Dispersion} \citep{Czech2019a} of a set of samples,
which detects branches or clades of the tree that exhibit a high heterogeneity across the samples
by visualizing a measure of dispersion (such as the variance) of the per-sample placement mass.
The method hence identifies branches and clades ``of interest'',
where samples differ in the amount of sequences being placed onto these parts of the tree.

The similarity between the placement distributions of two samples
can be measured with the \emph{phylogenetic Kantorovich-Rubinstein} (KR) distance \cite{Matsen2011a,Evans2012},
which is an adaptation of the Earth Mover's distance to phylogenetic placement.
The KR distance between two samples is a metric that quantifies
by {\em at least} how much the normalized mass distribution of one sample has to be moved across the reference tree
to obtain the distribution of the other sample.
In other words, it is the minimum work needed to solve the transportation problem between the two distributions (transforming one into the other),
and is related to the UniFrac distance \cite{Lozupone2005,Lozupone2007a}.
The distance is symmetrical, and increases the more mass needs to be moved
(that is, the more the abundances per branch and clade differ between the two samples),
and the larger the respective moving distance is (that is, the greater the phylogenetic distance along the branches of the tree between the clades is).
It is hence an intuitive and phylogenetically informed distance metric for placement data,
for example to quantify differences in the species composition of two environments.

\emph{Edge Principal Component Analysis} (Edge PCA) is a method to detect community structure,
which can also be employed for sample ordination and visualization \citep{Matsen2011a,Darling2014}.
Edge PCA identifies lineages of the RT that explain the greatest extent of variation between the sample communities,
and is computed via standard Principal Component Analysis on the per-edge imbalances across all samples.
The resulting principal components distinguish samples based on differences in abundances within clades of the reference tree.
See for example \figref{fig:Analyses:sub:EdgePCA},
where each point corresponds to a sample and is colorized according to a metadata variable of the sample,
showing that the ordination discriminates samples according to that variable.
Furthermore, as the eigenvectors of each principal component correspond to edges of the tree,
these can be visualized on the tree \citep{Matsen2011a,Czech2020a},
so that those edges and clades of the tree that explain differences between the samples can be identified,
e.\,g., with \toolname{guppy} \citep{Matsen2010} and \toolname{Archaeopteryx} \citep{Han2009},
or with \toolname{gappa} \citep{Czech2020}.
Principal components can also be computed from the balances instead of the imbalances \citep{Czech2020a}.


\paragraph{Clustering of Samples}
\label{sec:PlacementAnalysis:sub:SampleAnalysis:par:Clustering}

Given a measure of pairwise distance between samples, a fundamental task consists in clustering,
that is, finding groups of samples that are similar according to that measure.
\emph{Squash Clustering} \citep{Matsen2011a} is a hierarchical agglomerative clustering method for a set of placement samples, and is based on the KR distance.
Its results can be visualized as a clustering tree, where terminal nodes represent samples,
each inner node represents the cumulative distribution of all samples below that node (``squashed'' samples),
and distances along the tree edges are KR distances.
We show an example in \figref{fig:Analyses:sub:SquashClustering},
where each sample (terminal node) is colorized according to associated per-sample metadata variables (features measured for each sample), indicating that the clustering (based on the placement distribution) recovers characteristics of the samples based on that metadata variable.

The clustering hierarchy obtained from Squash Clustering grows with the number of samples,
which contains a lot of detail, but can be cumbersome to visualize and interpret for large datasets with many samples.
\emph{Phylogenetic $k$-means} clustering and \emph{Imbalance $k$-means} clustering \citep{Czech2019a} are further clustering approaches, which instead yield an assignment of each sample to one of a predefined number of $k$ clusters.
Phylogenetic $k$-means uses the KR distance for determining the cluster assignment of the samples, and hence yields results that are consistent with Squash Clustering, while Imbalance $k$-means uses edge imbalances, and hence is consistent with results obtained from Edge PCA.
Having the choice over the value $k$ can be beneficial to answer specific questions with a known set of categories of samples (e.\,g., different body locations where samples were obtained from), but is also considered a downside of $k$-means clustering.
Hence, various suggestions exist in the literature to select an appropriate $k$ that reflects the number of ``natural'' clusters in the data \cite{Thorndike1953,Rousseeuw1987,Bischof1999,Pelleg2000,Tibshirani2001,Hamerly2004}.
Visualizing the \emph{cluster centroids} obtained from both methods can further help to interpret results by showing the average distributions of all samples in one of the $k$ clusters; see again \citep{Czech2020a} for details.


\paragraph{Relationship with Environmental Metadata Variables} 
\label{sec:PlacementAnalysis:sub:SampleAnalysis:par:PostAnalysisMethods}

The above methods only implicitly take metadata into account,
e.\,g., by colorizing their resulting plots according to a variable.
Environmental variables can also be incorporated explicitly in phylogenetic placement analysis,
to more directly infer the relationships between the species composition of the samples
(e.\,g., in form of abundances per clade) and the environments these communities live in.

The \emph{Edge Correlation} \citep{Czech2019a} visualizes parts of the tree where species abundances
(as measured by the accumulated probability mass of each sample) exhibit a strong connection with a metadata variable,
see \figref{fig:Analyses:sub:Correlation}.
It is computed as the per-edge correlation coefficient between the per-sample metadata variable and
either the edge masses (highlighting individual edges), or imbalances or balances (highlighting clades) of each sample.

\emph{Placement-Factorization} \citep{Czech2019a,Czech2020a} is a more involved method.
It is an adaption of \emph{PhyloFactorization} \citep{Washburne2017a,Washburne2019} to phylogenetic placement data.
Its goal is to identify branches in the tree 
along which putative functional traits might have arisen in adaptation to changes in environmental variables.
In other words, it can detect clades of the reference tree whose abundances are linked to environmental factors.
By ``factoring out'' the clade with the strongest signal in each step of the algorithm (hence the name of the method),
nested dependencies with variables within clades can also be discovered,
see \figref{fig:Analyses:sub:PlacementFactorizationTree}.
This factorization of the tree into nested clades can further be used as an ordination tool to visualize
how samples are separated by changes along the factors, and as a dimensionality-reduction tool,
see \figref{fig:Analyses:sub:PlacementFactorizationOrdination}.
The method assesses the relationship between per-sample metadata features and the balances computed on the samples;
by using Generalized Linear Models, it allows to simultaneously incorporate multiple metadata variables of different types, such as numerical values (pH-value, temperature, latitude/longitude, etc),
binary values (presence/absence patterns, diseased or not),
or categorical values (body site that a sample was taken from).

\section{Conclusion and Outlook}
\label{sec:Conclusion}

In this review we broadly surveyed the concepts, methods, and software tools that constitute and relate to phylogenetic placement.
We have also presented guidelines and best practices for many typical use cases, showcased some common misconceptions and pitfalls, and introduced the most prominent downstream analysis methods.
Phylogenetic placement is a versatile approach that is particularly applicable in metagenomics (e.\,g., for metabarcoding data) and broader eDNA-based ecology studies.
It allows for the annotation of sequence data with phylogenetic information, and thereby to investigate the taxonomic content, functional capacity, diversity, and interactions of a community of organisms.
Further, it allows for comparing samples from multiple spatial and temporal locations, enabling the analysis of community patterns across time and space, as well as their association with environmental metadata variables.



Despite the growing popularity of phylogenetic placement, there are several methodological and usage aspects that will benefit from further developments.

Currently, significant effort is required to create high-quality reference trees.
We believe research effort should focus on simplifying this process, potentially through the design of methods that streamline and automate the commonly involved tasks.
For example, while there are some metrics that quantify the quality of an inferred phylogenetic tree \citep{Felsenstein1985,Dhar2016,Lemoine2018}, there is a lack of metrics to specifically evaluate the suitability of a tree for phylogenetic placement, given some expected input data.
Note that the \toolname{PEWO} testing framework \citep{Linard2020} (see \secref{sec:PhylogeneticPlacement:sub:RelatedTools}) represents a first step in this direction.

Ideally, reference trees and alignments should be created by, and shared in, research communities that investigate the same group(s) of organisms.
This would not only yield obtaining high-quality reference trees trivial, but would also immensely increase the comparability across studies, as well as their reproducibility.
Consequently, we would highly encourage such collaborations, and the public sharing of (perhaps even versioned instances of) gold-standard reference trees.
Notably, for some environments, first efforts into this direction have already been undertaken \cite{Berney2017,DelCampo2018,Rubinat-Ripoll2019,Rajter2021,Rajter2021a}.

Furthermore, as mentioned, there is a lack of established methods that evaluate placement quality in a standardized and meaningful way.
In particular, robust metrics are missing to distinguish the case where reference sequences of known species are missing from the tree from the case where the placed data actually contains yet undescribed species.
A classification based on the LWR and pendant length of the placement locations might offer a solution here.

Lastly, further work is required to connect environmental metadata to the results of phylogenetic placement.
Placement-based spatio-temporal methods are of high interest for addressing research questions in ecology and phylogeography.
For example, relating geo-locations of samples to their placement could indicate how species communities differ across space, while creating placement time series could show how community compositions develop and change over time.





\vspace{-0.5em}
\section*{Appendices}

%

\subsection*{Competing Interests}

The authors declare that they have no competing interests.

\subsection*{Author’s Contributions}

LC conceived the review and created the figures.
LC and PB drafted the manuscript.
All authors conducted literature research, and finalized and approved the manuscript.

\subsection*{Acknowledgments}
We wish to thank the Reviewers for the detailed and constructive comments that helped to improve this manuscript.
This work was financially supported by
the Carnegie Institution for Science at Stanford, California, USA,
the Klaus Tschira Stiftung gGmbH Foundation in Heidelberg, Germany,
and the Deutsche Forschungsgemeinschaft (grant DU1319/5-1).


\section*{References}
\bibliography{ms}

\begin{thebibliography}{100}

\bibitem{Edwards2013}
Edwards DJ, Holt KE (2013) {Beginner's guide to comparative bacterial genome
  analysis using next-generation sequence data.}
\newblock {\em Microbial Informatics and Experimentation} 3(1):2.

\bibitem{Sunagawa2013a}
Sunagawa S, et~al. (2013) {Metagenomic species profiling using universal
  phylogenetic marker genes}.
\newblock {\em Nature Methods} 10(12):1196.

\bibitem{Karsenti2011}
Karsenti E, et~al. (2011) {A holistic approach to marine Eco-systems biology}.
\newblock {\em PLoS Biology} 9(10):7--11.

\bibitem{Giner2016}
Giner CR, Forn I, Romac S, Logares R, Vargas CD (2016) {Environmental
  Sequencing Provides Reasonable Estimates of the Relative Abundance of
  Specific Picoeukaryotes}.
\newblock {\em Applied and Environmental Microbiology} 82(15):4757--4766.

\bibitem{LacoursireRoussel2016}
Lacoursi{\`{e}}re-Roussel A, C{\^{o}}t{\'{e}} G, Leclerc V, Bernatchez L (2016)
  Quantifying relative fish abundance with {eDNA}: a promising tool for
  fisheries management.
\newblock {\em Journal of Applied Ecology} 53(4):1148--1157.

\bibitem{Dupont2016}
Dupont A{\"{O}}C, Griffiths RI, Bell T, Bass D (2016) {Differences in soil
  micro-eukaryotic communities over soil pH gradients are strongly driven by
  parasites and saprotrophs}.
\newblock {\em Environmental Microbiology} 18(6):2010--2024.

\bibitem{Mahe2017}
Mah{\'{e}} F, et~al. (2017) {Parasites dominate hyperdiverse soil protist
  communities in Neotropical rainforests}.
\newblock {\em Nature Ecology {\&} Evolution} 1(4):0091.

\bibitem{Clare2022}
Clare EL, et~al. (2022) Measuring biodiversity from {DNA} in the air.
\newblock {\em Current Biology}.

\bibitem{Deiner2017}
Deiner K, et~al. (2017) Environmental {DNA} metabarcoding: Transforming how we
  survey animal and plant communities.
\newblock {\em Molecular Ecology} 26(21):5872--5895.

\bibitem{Ruppert2019}
Ruppert KM, Kline RJ, Rahman MS (2019) Past, present, and future perspectives
  of environmental dna (edna) metabarcoding: A systematic review in methods,
  monitoring, and applications of global edna.
\newblock {\em Global Ecology and Conservation} 17:e00547.

\bibitem{Huttenhower2012}
Huttenhower C, et~al. (2012) {Structure, function and diversity of the healthy
  human microbiome}.
\newblock {\em Nature} 486(7402):207--214.

\bibitem{Methe2012}
Meth{\'{e}} BA, et~al. (2012) {A framework for human microbiome research}.
\newblock {\em Nature} 486(7402):215--221.

\bibitem{Matsen2015}
Matsen FA (2015) {Phylogenetics and the Human Microbiome}.
\newblock {\em Systematic Biology} 64(1):e26--e41.

\bibitem{Wang2015}
Wang WL, et~al. (2015) {Application of metagenomics in the human gut
  microbiome}.
\newblock {\em World Journal of Gastroenterology} 21(3):803--814.

\bibitem{Hanson2016}
Hanson B, et~al. (2016) {Characterization of the bacterial and fungal
  microbiome in indoor dust and outdoor air samples: a pilot study}.
\newblock {\em Environmental Science: Processes {\&} Impacts} 18(6):713--724.

\bibitem{Gohli2019}
Gohli J, et~al. (2019) {The subway microbiome: Seasonal dynamics and direct
  comparison of air and surface bacterial communities}.
\newblock {\em Microbiome} 7(1):1--16.

\bibitem{Lorimer2019}
Lorimer J, et~al. (2019) {Making the microbiome public: Participatory
  experiments with DNA sequencing in domestic kitchens}.
\newblock {\em Transactions of the Institute of British Geographers}
  44(3):524--541.

\bibitem{ElRakaiby2019}
{El Rakaiby} MT, Gamal-Eldin S, Amin MA, Aziz RK (2019) {Hospital Microbiome
  Variations As Analyzed by High-Throughput Sequencing}.
\newblock {\em OMICS A Journal of Integrative Biology} 23(9):426--438.

\bibitem{Thomas2012}
Thomas T, Gilbert J, Meyer F (2012) {Metagenomics - a guide from sampling to
  data analysis}.
\newblock {\em Microbial Informatics and Experimentation} 2(1):3.

\bibitem{Oulas2015}
Oulas A, et~al. (2015) {Metagenomics: Tools and Insights for Analyzing
  Next-Generation Sequencing Data Derived from Biodiversity Studies}.
\newblock {\em Bioinformatics and Biology Insights} 9:75--88.

\bibitem{Escobar-Zepeda2015}
Escobar-Zepeda A, {Vera-Ponce De Le{\'{o}}n} A, Sanchez-Flores A (2015) {The
  road to metagenomics: From microbiology to DNA sequencing technologies and
  bioinformatics}.
\newblock {\em Frontiers in Genetics} 6(348):1--15.

\bibitem{Lindgreen2016}
Lindgreen S, Adair KL, Gardner PP (2016) {An evaluation of the accuracy and
  speed of metagenome analysis tools}.
\newblock {\em Scientific Reports} 6(1):19233.

\bibitem{Pettersson2009}
Pettersson E, Lundeberg J, Ahmadian A (2009) {Generations of Sequencing
  Technologies}.
\newblock {\em Genomics} 93(2):105--111.

\bibitem{Reuter2015}
Reuter JA, Spacek DV, Snyder MP (2015) {High-Throughput Sequencing
  Technologies}.
\newblock {\em Molecular Cell} 58(4):586--97.

\bibitem{Goodwin2016}
Goodwin S, McPherson JD, McCombie WR (2016) {Coming of age: ten years of
  next-generation sequencing technologies}.
\newblock {\em Nature Reviews Genetics} 17(6):333--351.

\bibitem{Logares2012}
Logares R, et~al. (2012) {Environmental microbiology through the lens of
  high-throughput DNA sequencing: Synopsis of current platforms and
  bioinformatics approaches}.
\newblock {\em Journal of Microbiological Methods} 91(1):106--113.

\bibitem{Mardis2013}
Mardis ER (2013) {Next-Generation Sequencing Platforms}.
\newblock {\em Annual Review of Analytical Chemistry} 6(1):287--303.

\bibitem{Pareek2011}
Pareek CS, Smoczynski R, Tretyn A (2011) {Sequencing technologies and genome
  sequencing}.
\newblock {\em Journal of Applied Genetics} 52(4):413--435.

\bibitem{Niedringhaus2011}
Niedringhaus TP, Milanova D, Kerby MB, Snyder MP, Barron AE (2011) {Landscape
  of Next-Generation Sequencing Technologies}.
\newblock {\em Analytical Chemistry} 83(12):4327--4341.

\bibitem{Mignardi2014}
Mignardi M, Nilsson M (2014) {Fourth-generation sequencing in the cell and the
  clinic}.
\newblock {\em Genome Medicine} 6(4):31.

\bibitem{Heather2016}
Heather JM, Chain B (2016) {The sequence of sequencers: The history of
  sequencing DNA}.
\newblock {\em Genomics} 107(1):1--8.

\bibitem{Mardis2017}
Mardis ER (2017) {DNA sequencing technologies: 2006-2016}.
\newblock {\em Nature protocols} 12(2):213--218.

\bibitem{Muir2016}
Muir P, et~al. (2016) {The real cost of sequencing: scaling computation to keep
  pace with data generation}.
\newblock {\em Genome Biology} 17(1):1--9.

\bibitem{Katz2022}
Katz K, et~al. (2022) {The Sequence Read Archive: a decade more of explosive
  growth}.
\newblock {\em Nucleic Acids Research} 50(D1):D387--D390.

\bibitem{Desai2012}
Desai N, Antonopoulos D, Gilbert JA, Glass EM, Meyer F (2012) {From genomics to
  metagenomics}.
\newblock {\em Current Opinion in Biotechnology} 23(1):72--76.

\bibitem{Temperton2012}
Temperton B, Giovannoni SJ (2012) {Metagenomics: Microbial diversity through a
  scratched lens}.
\newblock {\em Current Opinion in Microbiology} 15(5):605--612.

\bibitem{Peabody2015}
Peabody MA, {Van Rossum} T, Lo R, Brinkman FSL (2015) {Evaluation of shotgun
  metagenomics sequence classification methods using in silico and in vitro
  simulated communities.}
\newblock {\em BMC Bioinformatics} 16:363.

\bibitem{Koski2001}
Koski LB, Golding GB (2001) {The Closest BLAST Hit is Often not the Nearest
  Neighbor}.
\newblock {\em Journal of Molecular Evolution} 52(6):540--2.

\bibitem{Clemente2011}
Clemente JC, Jansson J, Valiente G (2011) {Flexible taxonomic assignment of
  ambiguous sequencing reads}.
\newblock {\em BMC Bioinformatics} 12(1):1--15.

\bibitem{Brady2009}
Brady A, Salzberg SL (2009) Phymm and {PhymmBL}: metagenomic phylogenetic
  classification with interpolated markov models.
\newblock {\em Nature Methods} 6(9):673--676.

\bibitem{Segata2012}
Segata N, et~al. (2012) Metagenomic microbial community profiling using unique
  clade-specific marker genes.
\newblock {\em Nature Methods} 9(8):811--814.

\bibitem{Truong2015}
Truong DT, et~al. (2015) {MetaPhlAn}2 for enhanced metagenomic taxonomic
  profiling.
\newblock {\em Nature Methods} 12(10):902--903.

\bibitem{Jamy2019a}
Jamy M, et~al. (2019) {Long metabarcoding of the eukaryotic rDNA operon to
  phylogenetically and taxonomically resolve environmental diversity}.
\newblock {\em Molecular Ecology Resources} 20(2):429--443.

\bibitem{Beghini2021}
Beghini F, et~al. (2021) Integrating taxonomic, functional, and strain-level
  profiling of diverse microbial communities with {bioBakery} 3.
\newblock {\em {eLife}} 10.

\bibitem{Ren2016}
Ren R, et~al. (2016) {Phylogenetic Resolution of Deep Eukaryotic and Fungal
  Relationships Using Highly Conserved Low-Copy Nuclear Genes}.
\newblock {\em Genome Biology and Evolution} 8(9):2683--701.

\bibitem{Hebert2003}
Hebert PDN, Cywinska A, Ball SL, DeWaard JR (2003) {Biological Identifications
  Through DNA Barcodes}.
\newblock {\em Proceedings in Biological Sciences} 270(1512):313--21.

\bibitem{Savolainen2005}
Savolainen V, Cowan RS, Vogler AP, Roderick GK, Lane R (2005) {Towards Writing
  the Encyclopedia of Life: An Introduction to DNA Barcoding}.
\newblock {\em Philosophical transactions of the Royal Society of London.
  Series B, Biological sciences} 360(1462):1805--11.

\bibitem{Kress2008}
Kress WJ, Erickson DL (2008) {DNA Barcodes: Genes, Genomics, and
  Bioinformatics}.
\newblock {\em Proceedings of the National Academy of Sciences of the United
  States of America} 105(8):2761--2.

\bibitem{Auladell2019}
Auladell A, S{\'{a}}nchez P, S{\'{a}}nchez O, Gasol JM, Ferrera I (2019)
  {Long-term seasonal and interannual variability of marine aerobic anoxygenic
  photoheterotrophic bacteria}.
\newblock {\em The ISME Journal 2019 13:8} 13(8):1975--1987.

\bibitem{Hleap2021}
Hleap JS, Littlefair JE, Steinke D, Hebert PD, Cristescu ME (2021) {Assessment
  of current taxonomic assignment strategies for metabarcoding eukaryotes}.
\newblock {\em Molecular Ecology Resources} 21(7):2190--2203.

\bibitem{Dunthorn2014}
Dunthorn M, et~al. (2014) {Placing environmental next-generation sequencing
  amplicons from microbial eukaryotes into a phylogenetic context}.
\newblock {\em Molecular Biology and Evolution} 31(4):993--1009.

\bibitem{Bass2018a}
Bass D, et~al. (2018) {Clarifying the Relationships between Microsporidia and
  Cryptomycota}.
\newblock {\em Journal of Eukaryotic Microbiology} 65(6):773--782.

\bibitem{Keck2018}
Keck F, Vasselon V, Rimet F, Bouchez A, Kahlert M (2018) {Boosting DNA
  metabarcoding for biomonitoring with phylogenetic estimation of operational
  taxonomic units' ecological profiles}.
\newblock {\em Molecular Ecology Resources} 18(6):1299--1309.

\bibitem{Muhlemann2020a}
M{\"{u}}hlemann B, et~al. (2020) {Diverse variola virus (smallpox) strains were
  widespread in northern Europe in the Viking Age}.
\newblock {\em Science} 369(6502).

\bibitem{Morel2020sars}
Morel B, et~al. (2020) {Phylogenetic Analysis of SARS-CoV-2 Data Is Difficult}.
\newblock {\em Molecular Biology and Evolution} 38(5):1777--1791.

\bibitem{Turakhia2021}
Turakhia Y, et~al. (2021) {Ultrafast Sample placement on Existing tRees (UShER)
  enables real-time phylogenetics for the SARS-CoV-2 pandemic}.
\newblock {\em Nature Genetics} 53(6):809--816.

\bibitem{Srinivasan2012}
Srinivasan S, et~al. (2012) {Bacterial communities in women with bacterial
  vaginosis: High resolution phylogenetic analyses reveal relationships of
  microbiota to clinical criteria}.
\newblock {\em PLOS ONE} 7(6):e37818.

\bibitem{Arenas2015}
Arenas M (2015) {Trends in substitution models of molecular evolution}.
\newblock {\em Frontiers in Genetics} 6(OCT):319.

\bibitem{Felsenstein2004}
Felsenstein J (2004) {\em {Inferring Phylogenies}}.
\newblock (Sinauer Associates Sunderland, MA), 2nd edition.

\bibitem{Yang2006}
Yang Z (2006) {\em {Computational Molecular Evolution}}.
\newblock (Oxford University Press).

\bibitem{Bininda-Emonds2001}
Bininda-Emonds OR, Brady SG, Kim J, Sanderson MJ (2001) {Scaling of accuracy in
  extremely large phylogenetic trees.}
\newblock {\em Pacific Symposium on Biocomputing} pp. 547--558.

\bibitem{Moret2002b}
Moret BM, Roshan U, Warnow T (2002) {Sequence-length requirements for
  phylogenetic methods} in {\em Lecture Notes in Computer Science}, eds.{}
  Guig{\'{o}} R, Gusfield D.
\newblock (Springer Berlin Heidelberg, Berlin, Heidelberg), Vol.{} 2452, pp.
  343--356.

\bibitem{VonMering2007}
von Mering C, et~al. (2007) {Quantitative Phylogenetic Assessment of Microbial
  Communities in Diverse Environments}.
\newblock {\em Science} 315(5815):1126--1130.

\bibitem{Matsen2010}
Matsen FA, Kodner RB, Armbrust EV (2010) {pplacer: linear time
  maximum-likelihood and Bayesian phylogenetic placement of sequences onto a
  fixed reference tree}.
\newblock {\em BMC Bioinformatics} 11(1):538.

\bibitem{Berger2011}
Berger S, Krompass D, Stamatakis A (2011) {Performance, accuracy, and web
  server for evolutionary placement of short sequence reads under maximum
  likelihood}.
\newblock {\em Systematic Biology} 60(3):291--302.

\bibitem{Czech2017}
Czech L, Huerta-Cepas J, Stamatakis A (2017) {A Critical Review on the Use of
  Support Values in Tree Viewers and Bioinformatics Toolkits}.
\newblock {\em Molecular Biology and Evolution} 17(4):383--384.

\bibitem{Matsen2012}
Matsen FA, Hoffman NG, Gallagher A, Stamatakis A (2012) {A format for
  phylogenetic placements}.
\newblock {\em PLoS ONE} 7(2):1--4.

\bibitem{JsonStandard}
Bray T (2014) {The JavaScript Object Notation (JSON) Data Interchange Format,
  RFC}.
\newblock Accessed: 2018-08-14.

\bibitem{JsonMemo}
Crockford D (2006) {The application/json Media Type for JavaScript Object
  Notation (JSON), RFC}.
\newblock Accessed: 2018-08-14.

\bibitem{Archie1986}
Archie J, et~al. (1986) {The Newick tree format}.

\bibitem{Czech2020}
Czech L, Barbera P, Stamatakis A (2020) {Genesis and Gappa: processing,
  analyzing and visualizing phylogenetic (placement) data}.
\newblock {\em Bioinformatics} 36(10):3263--3265.

\bibitem{bossa}
Lefeuvre P (2018) {\em BoSSA: A Bunch of Structure and Sequence Analysis}.
\newblock R package version 3.6.

\bibitem{Yu2017}
Yu G, Smith DK, Zhu H, Guan Y, Lam TTY (2017) {ggtree: An R package for
  visualization and annotation of phylogenetic trees with their covariates and
  other associated data}.
\newblock {\em Methods in Ecology and Evolution} 8(1):28--36.

\bibitem{Wang2020}
Wang LG, et~al. (2020) {Treeio: An R Package for Phylogenetic Tree Input and
  Output with Richly Annotated and Associated Data}.
\newblock {\em Molecular Biology and Evolution} 37(2):599--603.

\bibitem{Weisburg1991}
Weisburg WG, Barns SM, Pelletier DA, Lane DJ (1991) {16S Ribosomal DNA
  Amplification for Phylogenetic Study}.
\newblock {\em Journal of Bacteriology} 173(2):697--703.

\bibitem{Meyer2010}
Meyer A, Todt C, Mikkelsen NT, Lieb B (2010) {Fast evolving 18S rRNA sequences
  from Solenogastres (Mollusca) resist standard PCR amplification and give new
  insights into mollusk substitution rate heterogeneity}.
\newblock {\em BMC Evolutionary Biology 2010 10:1} 10(1):70.

\bibitem{Woese1977}
Woese CR, Fox GE (1977) {Phylogenetic structure of the prokaryotic domain: the
  primary kingdoms}.
\newblock {\em Proceedings of the National Academy of Sciences of the United
  States of America} 74(11):5088--90.

\bibitem{Woese1990}
Woese CR, Kandler O, Wheelis ML (1990) {Towards a natural system of organisms:
  proposal for the domains Archaea, Bacteria, and Eucarya}.
\newblock {\em Proceedings of the National Academy of Sciences of the United
  States of America} 87(12):4576--9.

\bibitem{Ji2013}
Ji Y, et~al. (2013) {Reliable, verifiable and efficient monitoring of
  biodiversity via metabarcoding}.
\newblock {\em Ecology Letters} 16(10):1245--1257.

\bibitem{Nguyen2014}
Nguyen NP, Mirarab S, Liu B, Pop M, Warnow T (2014) {TIPP: Taxonomic
  identification and phylogenetic profiling}.
\newblock {\em Bioinformatics} 30(24):3548--3555.

\bibitem{Hugerth2017}
Hugerth LW, Andersson AF (2017) {Analysing Microbial Community Composition
  through Amplicon Sequencing: From Sampling to Hypothesis Testing}.
\newblock {\em Frontiers in Microbiology} 8:1561.

\bibitem{Bartlett2003}
Bartlett JMS, Stirling D (2003) {\em {A Short History of the Polymerase Chain
  Reaction}}.
\newblock (PCR Protocols, Methods in Molecular Biology, volume 226), pp. 3--6.

\bibitem{Janssen2018}
Janssen S, et~al. (2018) {Phylogenetic Placement of Exact Amplicon Sequences
  Improves Associations with Clinical Information}.
\newblock {\em mSystems} 3(3):e00021--18.

\bibitem{Logares2014}
Logares R, et~al. (2014) {Metagenomic 16S rDNA Illumina tags are a powerful
  alternative to amplicon sequencing to explore diversity and structure of
  microbial communities}.
\newblock {\em Environmental Microbiology} 16(9):2659--2671.

\bibitem{Morgan2010}
Morgan JL, Darling AE, Eisen JA (2010) {Metagenomic sequencing of an in
  vitro-simulated microbial community}.
\newblock {\em PLoS ONE} 5(4):1--10.

\bibitem{Lee2009}
Lee ZMP, Bussema C, Schmidt TM (2009) {rrnDB}: documenting the number of {rRNA}
  and {tRNA} genes in bacteria and archaea.
\newblock {\em Nucleic Acids Research} 37(Database):D489--D493.

\bibitem{Kembel2012}
Kembel SW, Wu M, Eisen JA, Green JL (2012) Incorporating 16s gene copy number
  information improves estimates of microbial diversity and abundance.
\newblock {\em {PLoS} Computational Biology} 8(10):e1002743.

\bibitem{Angly2014}
Angly FE, et~al. (2014) {CopyRighter}: a rapid tool for improving the accuracy
  of microbial community profiles through lineage-specific gene copy number
  correction.
\newblock {\em Microbiome} 2(1).

\bibitem{PereiraFlores2019}
Pereira-Flores E, Gl\"{o}ckner FO, Fernandez-Guerra A (2019) Fast and accurate
  average genome size and 16s {rRNA} gene average copy number computation in
  metagenomic data.
\newblock {\em {BMC} Bioinformatics} 20(1).

\bibitem{Love2016}
Love MI, Hogenesch JB, Irizarry RA (2016) Modeling of {RNA}-seq fragment
  sequence bias reduces systematic errors in transcript abundance estimation.
\newblock {\em Nature Biotechnology} 34(12):1287--1291.

\bibitem{Silverman2021}
Silverman JD, et~al. (2021) Measuring and mitigating {PCR} bias in microbiota
  datasets.
\newblock {\em {PLOS} Computational Biology} 17(7):e1009113.

\bibitem{Eddy1995}
Eddy SR (1995) {Multiple alignment using hidden Markov models}.
\newblock {\em ISMB-95 Proceedings}.

\bibitem{Eddy1998}
Eddy SR (1998) {Profile hidden Markov models}.
\newblock {\em Bioinformatics} 14(9):755--763.

\bibitem{vanDijk2014}
van Dijk EL, Auger H, Jaszczyszyn Y, Thermes C (2014) Ten years of
  next-generation sequencing technology.
\newblock {\em Trends in Genetics} 30(9):418--426.

\bibitem{Piredda2021}
Piredda R, Grimm GW, Schulze ED, Denk T, Simeone MC (2021) {High-throughput
  sequencing of 5S-IGS in oaks: Exploring intragenomic variation and algorithms
  to recognize target species in pure and mixed samples}.
\newblock {\em Molecular Ecology Resources} 21(2):495--510.

\bibitem{Cardoni2022}
Cardoni S, et~al. (2022) {5S-IGS rDNA in wind-pollinated trees (Fagus L.)
  encapsulates 55 million years of reticulate evolution and hybrid origins of
  modern species}.
\newblock {\em The Plant Journal} 109(4):909--926.

\bibitem{Amarasinghe2020}
Amarasinghe SL, et~al. (2020) Opportunities and challenges in long-read
  sequencing data analysis.
\newblock {\em Genome Biology} 21(1).

\bibitem{Sharon2015}
Sharon I, et~al. (2015) Accurate, multi-kb reads resolve complex populations
  and detect rare microorganisms.
\newblock {\em Genome Research} 25(4):534--543.

\bibitem{Kuleshov2016}
Kuleshov V, et~al. (2016) Synthetic long-read sequencing reveals intraspecies
  diversity in the human microbiome.
\newblock {\em Nature Biotechnology} 34(1):64--69.

\bibitem{Ritter2020}
Ritter CD, et~al. (2020) {Advancing biodiversity assessments with environmental
  DNA: Long-read technologies help reveal the drivers of Amazonian fungal
  diversity}.
\newblock {\em Ecology and Evolution} 10(14):7509--7524.

\bibitem{Jeong2021}
Jeong J, et~al. (2021) The effect of taxonomic classification by full-length
  16s {rRNA} sequencing with a synthetic long-read technology.
\newblock {\em Scientific Reports} 11(1).

\bibitem{Blaxter2005}
Blaxter M, et~al. (2005) {Defining operational taxonomic units using DNA
  barcode data.}
\newblock {\em Philosophical Transactions of the Royal Society of London.
  Series B, Biological Sciences} 360(1462):1935--43.

\bibitem{Edgar2010}
Edgar RC (2010) {Search and clustering orders of magnitude faster than BLAST}.
\newblock {\em Bioinformatics} 26(19):2460--2461.

\bibitem{Fu2012}
Fu L, Niu B, Zhu Z, Wu S, Li W (2012) {CD}-{HIT}: accelerated for clustering
  the next-generation sequencing data.
\newblock {\em Bioinformatics} 28(23):3150--3152.

\bibitem{Rognes2016}
Rognes T, Flouri T, Nichols B, Quince C, Mah{\'{e}} F (2016) {VSEARCH: a
  versatile open source tool for metagenomics}.
\newblock {\em PeerJ} 4:e2584.

\bibitem{Westcott2015}
Westcott SL, Schloss PD (2015) {De novo clustering methods outperform
  reference-based methods for assigning 16S rRNA gene sequences to operational
  taxonomic units}.
\newblock {\em PeerJ} 2015(12).

\bibitem{Callahan2016}
Callahan BJ, et~al. (2016) {DADA}2: High-resolution sample inference from
  illumina amplicon data.
\newblock {\em Nature Methods} 13(7):581--583.

\bibitem{swarm3}
Mah{\'{e}} F, et~al. (2022) Swarm v3: towards tera-scale amplicon clustering.
\newblock {\em Bioinformatics} 38(1):267--269.

\bibitem{Zou2018}
Zou Q, Lin G, Jiang X, Liu X, Zeng X (2020) {Sequence clustering in
  bioinformatics: an empirical study}.
\newblock {\em Briefings in Bioinformatics} 21(1):1--10.

\bibitem{Hubert2014}
Hubert F, et~al. (2014) Multiple nuclear genes stabilize the phylogenetic
  backbone of the {genusQuercus}.
\newblock {\em Systematics and Biodiversity} 12(4):405--423.

\bibitem{Liede-Schumann2020}
Liede-Schumann S, et~al. (2020) {Phylogenetic relationships in the southern
  African genus Drosanthemum (Ruschioideae, Aizoaceae)}.
\newblock {\em PeerJ} 2020(3):e8999.

\bibitem{Berney2017}
Berney C, et~al. (2017) {UniEuk: Time to Speak a Common Language in
  Protistology!}
\newblock {\em Journal of Eukaryotic Microbiology} 38(1):42--49.

\bibitem{DelCampo2018}
del Campo J, et~al. (2018) {EukRef: Phylogenetic curation of ribosomal RNA to
  enhance understanding of eukaryotic diversity and distribution}.
\newblock {\em PLoS Biology} 16(9):1--14.

\bibitem{Rajter2021}
Rajter {\v{L}}, Ewers I, Graupner N, V{\v{d}}a{\v{c}}n{\'{y}} P, Dunthorn M
  (2021) {Colpodean ciliate phylogeny and reference alignments for phylogenetic
  placements}.
\newblock {\em European Journal of Protistology} 77.

\bibitem{Rajter2021a}
Rajter {\v{L}}, Dunthorn M (2021) {Ciliate SSU-rDNA reference alignments and
  trees for phylogenetic placements of metabarcoding data}.
\newblock {\em Metabarcoding and Metagenomics} 5:e69602.

\bibitem{Pruesse2007}
Pruesse E, et~al. (2007) {SILVA: a comprehensive online resource for quality
  checked and aligned ribosomal RNA sequence data compatible with ARB}.
\newblock {\em Nucleic Acids Research} 35(21):7188--7196.

\bibitem{Quast2013}
Quast C, et~al. (2013) {The SILVA ribosomal RNA gene database project: improved
  data processing and web-based tools}.
\newblock {\em Nucleic Acids Research} 41(D1):D590--D596.

\bibitem{Yilmaz2014}
Yilmaz P, et~al. (2014) {The SILVA and "All-species Living Tree Project (LTP)"
  taxonomic frameworks}.
\newblock {\em Nucleic Acids Research} 42(D1):D643--D648.

\bibitem{Sayers2009}
Sayers EW, et~al. (2009) {Database resources of the National Center for
  Biotechnology Information}.
\newblock {\em Nucleic Acids Research} 37(Database):D5--D15.

\bibitem{Benson2009}
Benson DA, Karsch-Mizrachi I, Lipman DJ, Ostell J, Sayers EW (2009) {GenBank}.
\newblock {\em Nucleic Acids Research} 37(Database):D26--D31.

\bibitem{DeSantis2006}
DeSantis TZ, et~al. (2006) {Greengenes, a chimera-checked 16S rRNA gene
  database and workbench compatible with ARB}.
\newblock {\em Applied and Environmental Microbiology} 72(7):5069--5072.

\bibitem{McDonald2012}
McDonald D, et~al. (2012) {An improved Greengenes taxonomy with explicit ranks
  for ecological and evolutionary analyses of bacteria and archaea}.
\newblock {\em ISME Journal} 6(3):610--618.

\bibitem{Cole2014}
Cole JR, et~al. (2014) {Ribosomal database project: data and tools for high
  throughput rRNA analysis}.
\newblock {\em Nucleic Acids Research} 42.

\bibitem{Wang2007}
Wang Q, Garrity GM, Tiedje JM, Cole JR (2007) {Na{\"{i}}ve Bayesian classifier
  for rapid assignment of rRNA sequences into the new bacterial taxonomy}.
\newblock {\em Applied and Environmental Microbiology} 73(16):5261--5267.

\bibitem{Balvociute2017}
Balvo{\v{c}}iūtė M, Huson DH (2017) {SILVA, RDP, Greengenes, NCBI and OTT ---
  how do these taxonomies compare?}
\newblock {\em BMC Genomics} 18(2):114.

\bibitem{Linard2020}
Linard B, Romashchenko N, Pardi F, Rivals E (2020) {PEWO: a collection of
  workflows to benchmark phylogenetic placement}.
\newblock {\em Bioinformatics}.

\bibitem{Czech2018}
Czech L, Barbera P, Stamatakis A (2018) {Methods for automatic reference trees
  and multilevel phylogenetic placement}.
\newblock {\em Bioinformatics} 35(7):1151--1158.

\bibitem{Notredame2000}
Notredame C, Higgins DG, Heringa J (2000) {T-coffee: a novel method for fast
  and accurate multiple sequence alignment}.
\newblock {\em Journal of Molecular Biology} 302(1):205--217.

\bibitem{Edgar2004}
Edgar RC (2004) {MUSCLE: multiple sequence alignment with high accuracy and
  high throughput}.
\newblock {\em Nucleic Acids Research} 32(5):1792--1797.

\bibitem{Katoh2002}
Katoh K, Misawa K, Kuma K, Miyata T (2002) {MAFFT: a novel method for rapid
  multiple sequence alignment based on fast Fourier transform}.
\newblock {\em Nucleic Acids Research} 30(14):3059--3066.

\bibitem{Kemena2009}
Kemena C, Notredame C (2009) {Upcoming challenges for multiple sequence
  alignment methods in the high-throughput era}.
\newblock {\em Bioinformatics} 25(19):2455--2465.

\bibitem{Pervez2014}
Pervez MT, et~al. (2014) {Evaluating the accuracy and efficiency of multiple
  sequence alignment methods.}
\newblock {\em Evolutionary bioinformatics online} 10:205--17.

\bibitem{Chatzou2016}
Chatzou M, et~al. (2016) {Multiple sequence alignment modeling: Methods and
  applications}.

\bibitem{Edgar2021}
Edgar RC (2021) {MUSCLE v5 enables improved estimates of phylogenetic tree
  confidence by ensemble bootstrapping}.
\newblock {\em bioRxiv} p. 2021.06.20.449169.

\bibitem{Kapli2020}
Kapli P, Yang Z, Telford MJ (2020) {Phylogenetic tree building in the genomic
  age}.
\newblock {\em Nature Reviews Genetics} 21(7):428--444.

\bibitem{Saitou1987}
Saitou N, Nei M (1987) {The Neighbor-Joining Method: A new Method for
  Reconstructing Phylogenetic Trees}.
\newblock {\em Molecular Biology and Evolution} 4(4):406--425.

\bibitem{Sankoff1975}
Sankoff D (1975) {Minimal Mutation Trees of Sequences}.
\newblock {\em SIAM Journal on Applied Mathematics} 28(1):35--42.

\bibitem{Holder2003}
Holder M, Lewis PO (2003) {Phylogeny estimation: traditional and Bayesian
  approaches.}
\newblock {\em Nature Reviews Genetics} 4(4):275--284.

\bibitem{Dhar2016}
Dhar A, Minin VN (2016) {Maximum Likelihood Phylogenetic Inference}.
\newblock {\em Encyclopedia of Evolutionary Biology} 2:499--506.

\bibitem{Nguyen2015a}
Nguyen LTT, Schmidt HA, {Von Haeseler} A, Minh BQ (2015) {IQ-TREE: A fast and
  effective stochastic algorithm for estimating maximum-likelihood
  phylogenies}.
\newblock {\em Molecular Biology and Evolution} 32(1):268--74.

\bibitem{Price2010}
Price MN, Dehal PS, Arkin AP (2010) {FastTree 2 – Approximately
  Maximum-Likelihood Trees for Large Alignments}.
\newblock {\em PLoS ONE} 5(3):e9490.

\bibitem{Stamatakis2014}
Stamatakis A (2014) {RAxML version 8: a tool for phylogenetic analysis and
  post-analysis of large phylogenies}.
\newblock {\em Bioinformatics} 30(9):1312--1313.

\bibitem{Kozlov2019a}
Kozlov AM, Darriba D, Flouri T, Morel B, Stamatakis A (2019) {RAxML-NG: A fast,
  scalable, and user-friendly tool for maximum likelihood phylogenetic
  inference}.
\newblock {\em Bioinformatics}.

\bibitem{Zhou2018}
Zhou X, Shen XX, Hittinger CT, Rokas A (2017) {Evaluating Fast Maximum
  Likelihood-Based Phylogenetic Programs Using Empirical Phylogenomic Data
  Sets}.
\newblock {\em Molecular Biology and Evolution} 35(2):486--503.

\bibitem{Huelsenbeck2001}
Huelsenbeck JP, Ronquist F, Nielsen R, Bollback JP (2001) {Bayesian inference
  of phylogeny and its impact on evolutionary biology.}
\newblock {\em Science (New York, N.Y.)} 294(5550):2310--4.

\bibitem{Ronquist2004}
Ronquist F (2004) {Bayesian inference of character evolution}.
\newblock {\em Trends in Ecology {\&} Evolution} 19(9):475--481.

\bibitem{Katoh2012}
Katoh K, Frith MC (2012) {Adding unaligned sequences into an existing alignment
  using MAFFT and LAST}.
\newblock {\em Bioinformatics} 28(23):3144--3146.

\bibitem{Berger2011a}
Berger S, Stamatakis A (2011) {Aligning short reads to reference alignments and
  trees}.
\newblock {\em Bioinformatics} 27(15):2068--2075.

\bibitem{Berger2012}
Berger S, Stamatakis A (2012) {PaPaRa 2.0: A Vectorized Algorithm for
  Probabilistic Phylogeny-Aware Alignment Extension}, (Heidelberg Institute for
  Theoretical Studies, Heidelberg), Technical report.

\bibitem{Loytynoja2012}
L{\"{o}}ytynoja A, Vilella AJ, Goldman N (2012) {Accurate extension of multiple
  sequence alignments using a phylogeny-aware graph algorithm}.
\newblock {\em Bioinformatics} 28(13):1684--1691.

\bibitem{Langmead2012}
Langmead B, Salzberg SL (2012) {Fast gapped-read alignment with Bowtie 2}.
\newblock {\em Nature Methods} 9(4):357--359.

\bibitem{Li2009}
Li H, Durbin R (2009) {Fast and accurate short read alignment with
  Burrows-Wheeler transform}.
\newblock {\em Bioinformatics} 25(14):1754--1760.

\bibitem{Li2010}
Li H, Durbin R (2010) {Fast and accurate long-read alignment with
  Burrows-Wheeler transform}.
\newblock {\em Bioinformatics} 26(5):589--595.

\bibitem{Stark2010}
Stark M, Berger SA, Stamatakis A, von Mering C (2010) {MLTreeMap - accurate
  Maximum Likelihood placement of environmental DNA sequences into taxonomic
  and functional reference phylogenies}.
\newblock {\em BMC Genomics} 11(1):461.

\bibitem{Strimmer2002}
Strimmer K, Rambaut A (2002) {Inferring confidence sets of possibly
  misspecified gene trees}.
\newblock {\em Proceedings of the Royal Society of London B: Biological
  Sciences} 269(1487):137--142.

\bibitem{Barbera2018}
Barbera P, et~al. (2018) {EPA-ng: Massively Parallel Evolutionary Placement of
  Genetic Sequences}.
\newblock {\em Systematic Biology}.

\bibitem{Linard2018}
Linard B, Swenson K, Pardi F (2019) {Rapid alignment-free phylogenetic
  identification of metagenomic sequences}.
\newblock {\em Bioinformatics} 35(18):328740.

\bibitem{Brown2012}
Brown DG, Truszkowski J (2012) {LSHPlace}: Fast phylogenetic placement using
  locality-sensitive hashing in {\em Biocomputing 2013}.
\newblock ({WORLD} {SCIENTIFIC}).

\bibitem{Balaban2019}
Balaban M, Sarmashghi S, Mirarab S (2019) {APPLES: Scalable Distance-Based
  Phylogenetic Placement with or without Alignments}.
\newblock {\em Systematic Biology}.

\bibitem{Dodsworth2015}
Dodsworth S (2015) Genome skimming for next-generation biodiversity analysis.
\newblock {\em Trends in Plant Science} 20(9):525--527.

\bibitem{Balaban2021}
Balaban M, Jiang Y, Roush D, Zhu Q, Mirarab S (2021) {Fast and accurate
  distance-based phylogenetic placement using divide and conquer}.
\newblock {\em Molecular Ecology Resources}.

\bibitem{Blanke2021}
Blanke M, Morgenstern B (2021) {App-SpaM: phylogenetic placement of short reads
  without sequence alignment}.
\newblock {\em Bioinformatics Advances} 1(1).
\newblock vbab027.

\bibitem{jukes1969evolution}
Jukes TH, Cantor CR (1969) {\em Mammalian protein metabolism}, ed.{} Munro H.
\newblock Vol.{}~3, pp. 21--132.

\bibitem{Felsenstein1978}
Felsenstein J (1978) {Cases in which Parsimony or Compatibility Methods will be
  Positively Misleading}.
\newblock {\em Systematic Biology} 27(4):401--410.

\bibitem{Bergsten2005}
Bergsten J (2005) A review of long-branch attraction.
\newblock {\em Cladistics} 21(2):163--193.

\bibitem{Degnan2009}
Degnan JH, Rosenberg NA (2009) Gene tree discordance, phylogenetic inference
  and the multispecies coalescent.
\newblock {\em Trends in Ecology {\&} Evolution} 24(6):332--340.

\bibitem{Rabiee2019}
Rabiee M, Mirarab S (2019) {INSTRAL}: Discordance-aware phylogenetic placement
  using quartet scores.
\newblock {\em Systematic Biology} 69(2):384--391.

\bibitem{Jiang2021}
Jiang Y, Balaban M, Zhu Q, Mirarab S (2021) {DEPP: Deep Learning Enables
  Extending Species Trees using Single Genes}.
\newblock {\em bioRxiv} p. 2021.01.22.427808.

\bibitem{Balaban2020}
Balaban M, Mirarab S (2020) Phylogenetic double placement of mixed samples.
\newblock {\em Bioinformatics} 36(Supplement{\_}1):i335--i343.

\bibitem{Hofreiter2001}
Hofreiter M, Serre D, Poinar HN, Kuch M, P{\"{a}}{\"{a}}bo S (2001) {Ancient
  DNA}.
\newblock {\em Nature Reviews Genetics} 2(5):353--359.

\bibitem{Martiniano2020}
Martiniano R, Sanctis BD, Hallast P, Durbin R (2020) {Placing ancient DNA
  sequences into reference phylogenies}.
\newblock {\em bioRxiv} p. 2020.12.19.423614.

\bibitem{Berger2010a}
Berger SA, Stamatakis A (2010) {Accuracy of morphology-based phylogenetic
  fossil placement under maximum likelihood}.
\newblock {\em 2010 ACS/IEEE International Conference on Computer Systems and
  Applications, AICCSA 2010}.

\bibitem{Bomfleur2015}
Bomfleur B, Grimm GW, McLoughlin S (2015) {Osmunda pulchella sp. nov. from the
  Jurassic of Sweden - reconciling molecular and fossil evidence in the
  phylogeny of modern royal ferns (Osmundaceae)}.
\newblock {\em BMC Evolutionary Biology} 15(1):1--25.

\bibitem{Zheng2018}
Zheng Q, Bartow-McKenney C, Meisel JS, Grice EA (2018) {HmmUFOtu}: An {HMM} and
  phylogenetic placement based ultra-fast taxonomic assignment and {OTU}
  picking tool for microbiome amplicon sequencing studies.
\newblock {\em Genome Biology} 19(1).

\bibitem{Carbone2016}
Carbone I, et~al. (2016) T-{BAS}: Tree-based alignment selector toolkit for
  phylogenetic-based placement, alignment downloads and metadata visualization:
  an example with the pezizomycotina tree of life.
\newblock {\em Bioinformatics} p. btw808.

\bibitem{Carbone2019}
Carbone I, et~al. (2019) T-{BAS} version 2.1: Tree-based alignment selector
  toolkit for evolutionary placement of {DNA} sequences and viewing alignments
  and specimen metadata on curated and custom trees.
\newblock {\em Microbiology Resource Announcements} 8(29).

\bibitem{Douglas2018}
Douglas GM, Beiko RG, Langille MG (2018) {Predicting the functional potential
  of the microbiome from marker genes using PICRUSt} in {\em Microbiome
  Analysis}.
\newblock (Springer), pp. 169--177.

\bibitem{Douglas2020a}
Douglas GM, et~al. (2020) {PICRUSt2 for prediction of metagenome functions}.
\newblock {\em Nature Biotechnology} pp. 1--5.

\bibitem{Erazo2021}
Erazo NG, Dutta A, Bowman JS (2021) From microbial community structure to
  metabolic inference using paprica.
\newblock {\em STAR Protocols} 2(4):101005.

\bibitem{Sempr2021}
Semp{\'{e}}r{\'{e}} G, et~al. (2021) {metaXplor}: an interactive viral and
  microbial metagenomic data manager.
\newblock {\em {GigaScience}} 10(2).

\bibitem{Mirarab2012}
Mirarab S, Nguyen N, Warnow T (2012) {SEPP: SAT{\'{e}}-Enabled Phylogenetic
  Placement} in {\em Pacific Symposium on Biocomputing}.
\newblock (World Scientific), pp. 247--258.

\bibitem{Koning2021}
Koning E, Phillips M, Warnow T (2021) {pplacerDC: a new scalable phylogenetic
  placement method} in {\em Proceedings of the 12th ACM Conference on
  Bioinformatics, Computational Biology, and Health Informatics}.
\newblock pp. 1--9.

\bibitem{Wedell2021}
Wedell E, Cai Y, Warnow T (2021) {Scalable and Accurate Phylogenetic Placement
  Using pplacer-XR} in {\em International Conference on Algorithms for
  Computational Biology}.
\newblock (Springer), pp. 94--105.

\bibitem{Liu2012}
Liu K, et~al. (2012) {SAT{\'{e}}-II: Very Fast and Accurate Simultaneous
  Estimation of Multiple Sequence Alignments and Phylogenetic Trees}.
\newblock {\em Systematic Biology} 61(1):90.

\bibitem{Pearson1988}
Pearson WR, Lipman DJ (1988) {Improved tools for biological sequence
  comparison}.
\newblock {\em Proceedings of the National Academy of Sciences}
  85(8):2444--2448.

\bibitem{Felsenstein1981}
Felsenstein J (1981) {Evolutionary trees from DNA sequences: A maximum
  likelihood approach}.
\newblock {\em Journal of Molecular Evolution} 17(6):368--376.

\bibitem{Letunic2016}
Letunic I, Bork P (2016) {Interactive tree of life (iTOL) v3: an online tool
  for the display and annotation of phylogenetic and other trees}.
\newblock {\em Nucleic Acids Research} 44(W1):W242--5.

\bibitem{Letunic2019}
Letunic I, Bork P (2019) {Interactive Tree of Life (iTOL) v4: Recent updates
  and new developments}.
\newblock {\em Nucleic Acids Research} 47(W1):W256--W259.

\bibitem{Han2009}
Han MV, Zmasek CM (2009) {phyloXML: XML for evolutionary biology and
  comparative genomics.}
\newblock {\em BMC Bioinformatics} 10:356.

\bibitem{Guillou2012}
Guillou L, et~al. (2012) {The Protist Ribosomal Reference database (PR2): a
  catalog of unicellular eukaryote small sub-unit rRNA sequences with curated
  taxonomy}.
\newblock {\em Nucleic Acids Research} 41(D1):D597--D604.

\bibitem{Ondov2011}
Ondov BD, Bergman NH, Phillippy AM (2011) {Interactive metagenomic
  visualization in a Web browser}.
\newblock {\em BMC Bioinformatics} 12(1):385.

\bibitem{Haas2011}
Haas BJ, et~al. (2011) {Chimeric 16S rRNA sequence formation and detection in
  Sanger and 454-pyrosequenced PCR amplicons}.
\newblock {\em Genome Research} 21(3):494.

\bibitem{Breitwieser2019}
Breitwieser FP, Lu J, Salzberg SL (2019) {A review of methods and databases for
  metagenomic classification and assembly}.
\newblock {\em Briefings in Bioinformatics} 20(4):1125--1136.

\bibitem{Altschul1990}
Altschul SF, Gish W, Miller W, Myers EW, Lipman DJ (1990) {Basic Local
  Alignment Search Tool}.
\newblock {\em Journal of Molecular Biology} 215(3):403--410.

\bibitem{Shah2018}
Shah N, Nute MG, Warnow T, Pop M (2018) {Misunderstood parameter of NCBI BLAST
  impacts the correctness of bioinformatics workflows}.
\newblock {\em Bioinformatics}.

\bibitem{Delsuc2020}
Delsuc F, Ranwez V (2020) {Accurate alignment of (meta)barcoding data sets
  using MACSE} in {\em {Phylogenetics in the Genomic Era}}, eds.{} Scornavacca
  C, Delsuc F, Galtier N.
\newblock ({No commercial publisher | Authors open access book}) No.{} 2.3, pp.
  2.3:1--2.3:31.

\bibitem{Krause2008}
Krause L, et~al. (2008) {Phylogenetic classification of short environmental DNA
  fragments}.
\newblock {\em Nucleic Acids Research} 36(7):2230--2239.

\bibitem{Schreiber2010}
Schreiber F, Gumrich P, Daniel R, Meinicke P (2010) {Treephyler: fast taxonomic
  profiling of metagenomes}.
\newblock {\em Bioinformatics} 26(7):960--961.

\bibitem{Boyd2018}
Boyd JA, Woodcroft BJ, Tyson GW (2018) {GraftM: a tool for scalable,
  phylogenetically informed classification of genes within metagenomes}.
\newblock {\em Nucleic acids research} 46(10):e59.

\bibitem{Caporaso2010a}
Caporaso JG, et~al. (2010) {QIIME allows analysis of high- throughput community
  sequencing data}.
\newblock {\em Nature Methods} 7(5):335--336.

\bibitem{Bolyen2019}
Bolyen E, et~al. (2019) {Reproducible, interactive, scalable and extensible
  microbiome data science using QIIME 2}.
\newblock {\em Nature Biotechnology} 37(8):852--857.

\bibitem{Schloss2009}
Schloss PD, et~al. (2009) {Introducing mothur: Open-source,
  platform-independent, community-supported software for describing and
  comparing microbial communities}.
\newblock {\em Applied and Environmental Microbiology} 75(23):7537--7541.

\bibitem{Lopez-Garcia2018}
L{\'{o}}pez-Garc{\'{i}}a A, et~al. (2018) {Comparison of mothur and QIIME for
  the analysis of rumen microbiota composition based on 16S rRNA amplicon
  sequences}.
\newblock {\em Frontiers in Microbiology} 9(DEC):1--11.

\bibitem{Prodan2020}
Prodan A, et~al. (2020) {Comparing bioinformatic pipelines for microbial 16S
  rRNA amplicon sequencing}.
\newblock {\em PLoS ONE} 15(1):1--19.

\bibitem{Huson2007a}
Huson DH, Auch AF, Qi J, Schuster SC (2007) {MEGAN analysis of metagenomic
  data}.
\newblock {\em Genome Research} 17(3):377--386.

\bibitem{Wood2014}
Wood DE, et~al. (2014) {Kraken: ultrafast metagenomic sequence classification
  using exact alignments}.
\newblock {\em Genome Biology} 15(3):R46.

\bibitem{Wood2019}
Wood DE, Lu J, Langmead B (2019) {Improved metagenomic analysis with Kraken 2}.
\newblock {\em Genome Biology} 20(1):1--13.

\bibitem{Menzel2016a}
Menzel P, Ng KL, Krogh A (2016) {Fast and sensitive taxonomic classification
  for metagenomics with Kaiju}.
\newblock {\em Nature Communications} 7(1):1--9.

\bibitem{Sczyrba2017}
Sczyrba A, et~al. (2017) {Critical Assessment of Metagenome Interpretation—a
  benchmark of metagenomics software}.
\newblock {\em Nature Methods} 14(11):1063--1071.

\bibitem{Bremges2018}
Bremges A, McHardy AC (2018) {Critical Assessment of Metagenome Interpretation
  Enters the Second Round}.
\newblock {\em mSystems} 3(4).

\bibitem{Meyer2019}
Meyer F, et~al. (2019) {Assessing taxonomic metagenome profilers with OPAL}.
\newblock {\em Genome Biology} 20(1):51.

\bibitem{Ye2019}
Ye SH, Siddle KJ, Park DJ, Sabeti PC (2019) {Benchmarking Metagenomics Tools
  for Taxonomic Classification}.
\newblock {\em Cell} 178(4):779--794.

\bibitem{Smith2017}
Smith SA, Pease JB (2017) {Heterogeneous molecular processes among the causes
  of how sequence similarity scores can fail to recapitulate phylogeny}.
\newblock {\em Briefings in Bioinformatics} 18(3):451--457.

\bibitem{Darling2014}
Darling AE, et~al. (2014) {PhyloSift: phylogenetic analysis of genomes and
  metagenomes.}
\newblock {\em PeerJ} 2:e243.

\bibitem{Matsen2012a}
Matsen Fa, Gallagher A (2012) {Reconciling taxonomy and phylogenetic inference:
  formalism and algorithms for describing discord and inferring taxonomic
  roots.}
\newblock {\em Algorithms for molecular biology : AMB} 7(1):8.

\bibitem{Kozlov2016}
Kozlov AM, Zhang J, Yilmaz P, Gl{\"{o}}ckner FO, Stamatakis A (2016)
  {Phylogeny-aware identification and correction of taxonomically mislabeled
  sequences}.
\newblock {\em Nucleic Acids Research} 44(11):5022--5033.

\bibitem{Petrenko2015}
Petrenko P, Lobb B, Kurtz DA, Neufeld JD, Doxey AC (2015) {MetAnnotate:
  Function-specific taxonomic profiling and comparison of metagenomes}.
\newblock {\em BMC Biology} 13(1):92.

\bibitem{Shah2021}
Shah N, Molloy EK, Pop M, Warnow T (2021) {TIPP2: metagenomic taxonomic
  profiling using phylogenetic markers}.
\newblock {\em Bioinformatics}.

\bibitem{Morgan-Lang2020}
Morgan-Lang C, et~al. (2020) {TreeSAPP: the Tree-based Sensitive and Accurate
  Phylogenetic Profiler}.
\newblock {\em Bioinformatics} 36(18):4706--4713.

\bibitem{Schon2019}
Sch{\"{o}}n ME, Eme L, Ettema TJG (2019) {PhyloMagnet: fast and accurate
  screening of short-read meta-omics data using gene-centric phylogenetics}.
\newblock {\em Bioinformatics} 36(6):1718--1724.

\bibitem{Tucker2017guide}
Tucker CM, et~al. (2017) A guide to phylogenetic metrics for conservation,
  community ecology and macroecology.
\newblock {\em Biological Reviews} 92(2):698--715.

\bibitem{Faith1992}
Faith DP (1992) Conservation evaluation and phylogenetic diversity.
\newblock {\em Biological Conservation} 61(1):1--10.

\bibitem{McCoy2013}
McCoy CO, Matsen FA (2013) Abundance-weighted phylogenetic diversity measures
  distinguish microbial community states and are robust to sampling depth.
\newblock {\em {PeerJ}} 1:e157.

\bibitem{Barbera2020a}
Barbera P, Czech L, Lutteropp S, Stamatakis A (2020) {SCRAPP: A tool to assess
  the diversity of microbial samples from phylogenetic placements}.
\newblock {\em Molecular Ecology Resources} 21(1):1755--0998.13255.

\bibitem{Zhang2013}
Zhang J, Kapli P, Pavlidis P, Stamatakis A (2013) {A general species
  delimitation method with applications to phylogenetic placements}.
\newblock {\em Bioinformatics} 29(22):2869--2876.

\bibitem{Kapli2017}
Kapli P, et~al. (2017) {Multi-rate Poisson tree processes for single-locus
  species delimitation under maximum likelihood and Markov chain Monte Carlo}.
\newblock {\em Bioinformatics} 33(11):1630--1638.

\bibitem{Agapow2004}
Agapow PM, et~al. (2004) The impact of species concept on biodiversity studies.
\newblock {\em The Quarterly Review of Biology} 79(2):161--179.

\bibitem{Lozupone2005}
Lozupone C, Knight R (2005) {UniFrac: a New Phylogenetic Method for Comparing
  Microbial Communities}.
\newblock {\em Applied and Environmental Microbiology} 71(12):8228--8235.

\bibitem{Lozupone2007a}
Lozupone CA, Hamady M, Kelley ST, Knight R (2007) {Quantitative and Qualitative
  $\beta$ Diversity Measures Lead to Different Insights into Factors That
  Structure Microbial Communities}.
\newblock {\em Applied and Environmental Microbiology} 73(5):1576--1585.

\bibitem{Evans2012}
Evans SN, Matsen FA (2012) {The phylogenetic Kantorovich-Rubinstein metric for
  environmental sequence samples}.
\newblock {\em Journal of the Royal Statistical Society. Series B: Statistical
  Methodology} 74:569--592.

\bibitem{Nugent1991}
Nugent RP, Krohn MA, Hillier SL (1991) {Reliability of diagnosing bacterial
  vaginosis is improved by a standardized method of gram stain interpretation.}
\newblock {\em Journal of Clinical Microbiology} 29(2):297--301.

\bibitem{Czech2019a}
Czech L, Stamatakis A (2019) {Scalable methods for analyzing and visualizing
  phylogenetic placement of metagenomic samples}.
\newblock {\em PLOS ONE} 14(5):e0217050.

\bibitem{Matsen2011a}
Matsen FA, Evans SN (2011) {Edge principal components and squash clustering:
  using the special structure of phylogenetic placement data for sample
  comparison.}
\newblock {\em PLOS ONE} 8(3):1--17.

\bibitem{Czech2020a}
Czech L (2020) {Novel Methods for Analyzing and Visualizing Phylogenetic
  Placements. Ph.D. thesis.}
\newblock {\em Karlsruher Institut für Technologie,} Karlsruhe, Germany.

\bibitem{Li2015a}
Li H (2015) {Microbiome, Metagenomics, and High-Dimensional Compositional Data
  Analysis}.
\newblock {\em Annual Review of Statistics and Its Application} 2(1):73--94.

\bibitem{Gloor2017}
Gloor GB, Macklaim JM, Pawlowsky-Glahn V, Egozcue JJ (2017) {Microbiome
  Datasets Are Compositional: And This Is Not Optional}.
\newblock {\em Frontiers in Microbiology} 8:2224.

\bibitem{Quinn2018}
Quinn TP, Erb I, Richardson MF, Crowley TM (2018) {Understanding sequencing
  data as compositions: an outlook and review}.
\newblock {\em Bioinformatics} 34(16):2870--2878.

\bibitem{Silverman2017}
Silverman JD, Washburne AD, Mukherjee S, David LA (2017) {A phylogenetic
  transform enhances analysis of compositional microbiota data}.
\newblock {\em eLife} 6:e21887.

\bibitem{Calle2019}
Calle ML (2019) {Statistical Analysis of Metagenomics Data}.
\newblock {\em Genomics {\&} Informatics} 17(1):e6.

\bibitem{Kanagawa2003}
Kanagawa T (2003) {Bias and Artifacts in Multitemplate Polymerase Chain
  Reactions (PCR)}.
\newblock {\em Journal of Bioscience and Bioengineering} 96(4):317--323.

\bibitem{Weiss2017}
Weiss S, et~al. (2017) {Normalization and microbial differential abundance
  strategies depend upon data characteristics}.
\newblock {\em Microbiome} 5(1):27.

\bibitem{Aitchison1986}
Aitchison J (1986) {\em {The statistical analysis of compositional data}}.
\newblock (Chapman and Hall London).

\bibitem{Jackson1997}
Jackson DA (1997) {Compositional data in community ecology: The paradigm or
  peril of proportions?}
\newblock {\em Ecology} 78(3):929--940.

\bibitem{Tsilimigras2016}
Tsilimigras MCB, Fodor AA (2016) {Compositional data analysis of the
  microbiome: fundamentals, tools, and challenges}.
\newblock {\em Annals of Epidemiology} 26(5):330--335.

\bibitem{Gloor2016a}
Gloor GB, Macklaim JM, Vu M, Fernandes AD (2016) {Compositional uncertainty
  should not be ignored in high-throughput sequencing data analysis}.
\newblock {\em Austrian Journal of Statistics} 45(4):73.

\bibitem{Gotelli2001}
Gotelli NJ, Colwell RK (2001) {Quantifying biodiversity: procedures and
  pitfalls in the measurement and comparison of species richness}.
\newblock {\em Ecology Letters} 4(4):379--391.

\bibitem{McMurdie2014}
McMurdie PJ, Holmes S (2014) {Waste Not, Want Not: Why Rarefying Microbiome
  Data Is Inadmissible}.
\newblock {\em PLoS Computational Biology} 10(4):e1003531.

\bibitem{Egozcue2005}
Egozcue JJ, Pawlowsky-Glahn V (2005) {Groups of Parts and Their Balances in
  Compositional Data Analysis}.
\newblock {\em Mathematical Geology} 37(7):795--828.

\bibitem{Pawlowsky-Glahn2015}
Pawlowsky-Glahn V, Egozcue JJ, Tolosana-Delgado R (2015) {\em {Modeling and
  Analysis of Compositional Data}}.
\newblock (John Wiley {\&} Sons, Chichester, UK), p. 272.

\bibitem{Egozcue2003}
Egozcue JJ, Pawlowsky-Glahn V, Mateu-Figueras G, Barcel{\'{o}}-Vidal C (2003)
  {Isometric Logratio Transformations for Compositional Data Analysis}.
\newblock {\em Mathematical Geology} 35(3):279--300.

\bibitem{Tyson2004}
Tyson GW, et~al. (2004) {Community structure and metabolism through
  reconstruction of microbial genomes from the environment}.
\newblock {\em Nature} 428(6978):37--43.

\bibitem{Thorndike1953}
Thorndike RL (1953) {Who belongs in the family?}
\newblock {\em Psychometrika} 18(4):267--276.

\bibitem{Rousseeuw1987}
Rousseeuw PJ (1987) {Silhouettes: A graphical aid to the interpretation and
  validation of cluster analysis}.
\newblock {\em Journal of Computational and Applied Mathematics} 20:53--65.

\bibitem{Bischof1999}
Bischof H, Leonardis A, Selb A (1999) {MDL Principle for Robust Vector
  Quantisation}.
\newblock {\em Pattern Analysis {\&} Applications} 2(1):59--72.

\bibitem{Pelleg2000}
Pelleg D, Moore AW (2000) {X-means: Extending K-means with Efficient Estimation
  of the Number of Clusters.} in {\em ICML}.
\newblock Vol.{}~1, pp. 727--734.

\bibitem{Tibshirani2001}
Tibshirani R, Walther G, Hastie T (2001) {Estimating the number of clusters in
  a data set via the gap statistic}.
\newblock {\em Journal of the Royal Statistical Society: Series B (Statistical
  Methodology)} 63(2):411--423.

\bibitem{Hamerly2004}
Hamerly G, Elkan C (2004) {Learning the k in k-means} in {\em Advances in
  Neural Information Processing Systems 16}, eds.{} Thrun S, Saul LK,
  Sch{\"{o}}lkopf PB.
\newblock (MIT Press), pp. 281--288.

\bibitem{Washburne2017a}
Washburne AD, et~al. (2017) {Phylogenetic factorization of compositional data
  yields lineage-level associations in microbiome datasets}.
\newblock {\em PeerJ} 5:e2969.

\bibitem{Washburne2019}
Washburne AD, et~al. (2019) {Phylofactorization: a graph partitioning algorithm
  to identify phylogenetic scales of ecological data}.
\newblock {\em Ecological Monographs} 89(2):e01353.

\bibitem{Felsenstein1985}
Felsenstein J (1985) {Confidence} {limits} {on} {phylogenies}: {an} {approach}
  {using} {the} {bootstrap}.
\newblock {\em Evolution} 39(4):783--791.

\bibitem{Lemoine2018}
Lemoine F, et~al. (2018) Renewing felsenstein's phylogenetic bootstrap in the
  era of big data.
\newblock {\em Nature} 556(7702):452--456.

\bibitem{Rubinat-Ripoll2019}
Rubinat-Ripoll L (2019) lrubinat/photoreft: a 16s rdna reference tree
  representing the main groups of picophototrophic eukaryotes and prokaryotes.

\end{thebibliography}

\end{document}